\documentclass[preprint,3p,11pt]{elsarticle}
\usepackage{amssymb}
\usepackage{color}
\usepackage{amsmath}
\usepackage{xcolor}
\usepackage[linesnumbered,vlined,ruled]{algorithm2e}
\usepackage{tcolorbox}
\usepackage{amsthm}
\usepackage{pifont}
\usepackage{wrapfig}
\usepackage{subfigure}
\usepackage{multirow}
\usepackage{array}
\usepackage{booktabs}
\usepackage[figuresright]{rotating}
\usepackage[symbol]{footmisc}

\usepackage{colortbl}
\definecolor{lightgrayfortable}{gray}{0.92}

\usepackage{footnote}
\makesavenoteenv{tabular}

\usepackage{makecell}

\usepackage{afterpage}

\usepackage{tabularx}
\usepackage{pdflscape}

\usepackage{longtable}
\usepackage{wasysym} % \LEFTcircle

\usepackage{url}
\newcommand{\etc}{\hbox{\emph{etc.}}\xspace}
\newcommand{\eg}{\hbox{\emph{e.g.}}\xspace}
\newcommand{\ie}{\hbox{\emph{i.e.}}\xspace}

\newcommand{\etal}{\hbox{\emph{et al.}}\xspace}

\def\Figref#1{Fig.~\ref{#1}}
\def\Secref#1{\S~\ref{#1}}
\def\Eqref#1{Eq.~(\ref{#1})}
\def\Defref#1{\textbf{Def}.~\ref{#1}}
\def\Tabref#1{Table~\ref{#1}}

\newcommand{\tabincell}[2]{\begin{tabular}{@{}#1@{}}#2\end{tabular}}
\newcommand{\vleftthead}[1]{\makebox[1em][l]{\rotatebox{270}{{#1}}}}
\newcommand{\vlefttheadC}[1]{\makebox[1em][c]{\rotatebox{270}{{#1}}}}

\theoremstyle{definition}
\newtheorem{definition}{Def.}

\usepackage{tikz}
\newcommand*\emptycircle[1][1.0ex]{\tikz\draw (0,0) circle (#1);} 
\newcommand*\halfcircle[1][1.0ex]{%
  \begin{tikzpicture}
  \draw[fill] (0,0)-- (90:#1) arc (90:270:#1) -- cycle ;
  \draw (0,0) circle (#1);
  \end{tikzpicture}}
\newcommand*\fullcircle[1][1.0ex]{\tikz\fill (0,0) circle (#1);} 

\journal{Computer \& Security}
\usepackage{fancyhdr}
\begin{document}

\begin{frontmatter}


% \author{Name\corref{cor1}\fnref{label2}}
% \affiliation{organization={},

\title{\Large \textbf{Adversarial Attacks against Windows PE Malware Detection: A Survey of the State-of-the-Art}}

\author[iscas]{Xiang~Ling}\ead{lingxiang@iscas.ac.cn}
\author[jd]{Lingfei~Wu}\ead{lwu@email.wm.edu}
\author[zju]{Jiangyu~Zhang}\ead{zhangjiangyu@zju.edu.cn}
\author[zju]{Zhenqing~Qu}\ead{quzhenqing@zju.edu.cn}
\author[zju]{Wei~Deng}\ead{dengwei@zju.edu.cn}
\author[zju]{Xiang~Chen}\ead{wasdnsxchen@gmail.com}
\author[zust]{Yaguan~Qian}\ead{qianyaguan@zust.edu.cn}
\author[zju]{Chunming Wu}\ead{wuchunming@zju.edu.cn}
\author[zju]{Shouling Ji}\ead{sji@zju.edu.cn}
\author[iscas]{Tianyue Luo}\ead{tianyue@iscas.ac.cn}
\author[iscas]{Jingzheng Wu\corref{cor1}}\ead{jingzheng08@iscas.ac.cn}
\author[iscas]{Yanjun Wu\corref{cor1}}\ead{yanjun@iscas.ac.cn}

\cortext[cor1]{Corresponding Author.}

\affiliation[iscas]{organization={Institute of Software, Chinese Academy of Sciences},
            city={Beijing},
            postcode={100190}, 
            state={Beijing},
            country={China}}

\affiliation[jd]{organization={Pinterest},
            city={New York},
            postcode={10018}, 
            state={NY},
            country={USA}}

\affiliation[zust]{organization={Zhejiang University of Science and Technology},
            city={Hangzhou},
            postcode={310023}, 
            state={Zhejiang},
            country={China}}
            
\affiliation[zju]{organization={Zhejiang University},
            city={Hangzhou},
            postcode={310027}, 
            state={Zhejiang},
            country={China}}

\begin{abstract}
Malware has been one of the most damaging threats to computers that span across multiple operating systems and various file formats.
To defend against ever-increasing and ever-evolving malware, tremendous efforts have been made to propose a variety of malware detection that attempt to effectively and efficiently detect malware so as to mitigate possible damages as early as possible.
Recent studies have shown that, on the one hand, existing machine learning (ML) and deep learning (DL) techniques enable superior solutions in detecting newly emerging and previously unseen malware.
% provide superior detection solutions in detecting new and previously unseen malwares.
However, on the other hand, ML and DL models are inherently vulnerable to adversarial attacks in the form of adversarial examples, which are maliciously generated by slightly and carefully perturbing the legitimate inputs to misbehave.
Adversarial attacks are initially studied in the domain of computer vision like image classification, and then quickly extended to other domains, including natural language processing, audio recognition, and even malware detection.

In this paper, we focus on malware with the file format of portable executable (PE) in the family of Windows operating systems, namely \textbf{Windows PE malware}, as a representative case to study the adversarial attack methods in such adversarial settings.
To be specific, we start by first outlining the general learning framework of Windows PE malware detection based on ML/DL and subsequently highlighting three unique challenges of performing adversarial attacks in the context of Windows PE malware.
Then, we conduct a comprehensive and systematic review to categorize the state-of-the-art adversarial attacks against PE malware detection, as well as corresponding defenses to increase the robustness of Windows PE malware detection.
Finally, we conclude the paper by first presenting other related attacks against Windows PE malware detection beyond the adversarial attacks and then shedding light on future research directions and opportunities.
In addition, a curated resource list of adversarial attacks and defenses for Windows PE malware detection is also available at \url{https://github.com/ryderling/adversarial-attacks-and-defenses-for-windows-pe-malware-detection}.
\end{abstract}

\begin{keyword}
Portable Executable \sep Malware Detection \sep Machine Learning \sep Adversarial Machine Learning \sep Deep Learning \sep Adversarial Attack
\end{keyword}

\end{frontmatter}



\section{Introduction}
With the rapid development and advancement of information technology, computer systems are playing an indispensable and ubiquitous role in our daily lives.
Meanwhile, the cyberattack that attempts to maliciously exploit the computer system with malicious intentions (\eg, damaging computers or gaining economic profits) has been an important type of ever-increasing and constantly evolving security threat in our society.
Malware (\ie, short for \textbf{Mal}icious soft\textbf{ware}) is one of the most common and powerful cyberattacks for attackers to perform malicious activities in computer systems, such as stealing confidential information without permissions, compromising the whole system, and demanding for a large ransom.
While malware spans across multiple operating systems (\eg, Windows, Linux, macOS, Android, \etc) with various file formats, such as portable executable (PE), executable and linkable format (ELF), Mach-O, Android application package (APK), and portable document format (PDF), we focus on malware with the PE files in the family of Windows operating systems (namely \textbf{Windows PE malware}) in this paper due to the following two reasons.
First, malware analysis techniques (\eg, detection methods) for Windows PE files are mostly different from those for other files like APK and PDF files because their underlying operating systems, the file format, and execution modes are significantly different from each other.
Research shows there is no universal malware detection that can satisfactorily detect all kinds of malware, and thus existing literature papers on malware analysis commonly point out what specific operating system they target and what file format they are~\cite{ucci2019survey,raff2020survey}.
That is the very first and most important reason why we focus on Windows PE malware in this paper.
Second, Windows is the most worldwide popular and long-standing operation system for end users while the malware in the file format of PE constitutes the earliest and maximum studied threat in the wild~\cite{schultz2001data}.
According to the statistics of Kaspersky Lab at the end of 2020, there are an average of 360,000 malware detected by Kaspersky per day and over 90\% of which are Windows PE malware~\cite{Kaspersky_2020}.
Similar statistical trends have been reported by Kaspersky Lab at the end of 2021~\cite{Kaspersky_2021}, indicating Windows PE files are still not sufficiently protected until now.

To mitigate and address the ever-increasing number of security threats caused by Windows PE malware, there are tremendous research efforts have been made to detect Windows PE malware effectively and efficiently~\cite{idika2007survey,ye2017survey,ucci2019survey,raff2020survey,ceschin2020machine}.
In particular, traditional malware detection can be traced back to signature-based malware detection, which determines whether a given suspicious software is malicious or not (\ie, malware or goodware) by comparing its signature with all signatures from the maintained database of malware that has been previously collected and confirmed.
It is obvious that the fatal flaw of signature-based malware detection is that it can only detect previously collected and known malware due to the heavy reliance on the malware database.
In the last few decades, inspired by the great successes of ML and DL in various long-standing real-world tasks (\eg, computer vision, natural language processing, speech recognition, \etc), a variety of ML/DL-based malware detection methods~\cite{ye2017survey,ucci2019survey,raff2020survey,ceschin2020machine} that leverage the high capacity of ML/DL models have been adapted and presented to detect Windows PE malware.
In general, these ML/DL-based malware detection methods claim that they can generalize well to predict the new and previously unseen (\ie, zero-day) malware instances due to the inherent generalizability of ML/DL models.

Unfortunately, recent studies have demonstrated that existing ML/DL models are inherently vulnerable to adversarial attacks in the form of adversarial examples, which are maliciously generated by slightly and carefully perturbing the legitimate inputs to confuse the target ML/DL models~\cite{goodfellow2014explaining,carlini2017towards}.
Since the creation of adversarial attacks, most research papers focused on studying adversarial attacks in the domain of computer vision~\cite{akhtar2018threat}, \eg, slightly and carefully perturbing a ``Persian cat'' image $x$ such that the resulting adversarial image $x'$ can be misclassified as a ``lionfish'' by the target image classifier.
Normally, in the context of images, most proposed adversarial attacks resolve to the feature-space attack like various gradient-based methods, which can be directly applied to generate adversarial images.
Until now, there have been a large number of adversarial attack methods and corresponding defense mechanisms being proposed by security researchers and practitioners in both academia and industry~\cite{chakraborty2018adversarial,zhang2019adversarial,serban2020adversarial,machado2021adversarial,long2022survey}.

Inspired by those studies of adversarial attacks in the context of images, a natural question arises is that, \textbf{is it feasible to perform adversarial attacks against existing malware detection methods, especially against ML/DL based PE malware detection?}
To answer the aforementioned question, in recent five years, security researchers and practitioners have proposed lots of adversarial attacks in the context of malware~\cite{kreuk2018deceiving,suciu2019exploring,chen2019adversarial,qiao2022adversarial,al2018adversarial,liu2019atmpa,khormali2019copycat,park2019generation,li2020adversarial,kolosnjaji2018adversarial,demetrio2019explaining,demetrio2020adversarial,lucas2021malware,hu2017generating,kawai2019improved,chen2017adversarial,hu2017black,rosenberg2018generic,fadadu2019evading,rosenberg2020query,rosenberg2020generating,zhang2020semantic,anderson2017evading,wu2018enhancing,chen2020generating,fang2019evading,fang2020deepdetectnet,ebrahimi2021binary,labaca2021aimed,li2021irl,castro2019armed,ceschin2019shallow,song2020automatic,castro2019aimed,wang2020mdea,demetrio2020efficient,yuan20black,zhong2020malfox,fleshman2018static}, requiring that the generated adversarial malware file should not only be misclassified as the ``goodware'' by the target malware detection model, but also behaves exactly the same as the original malware file.
Compared to the adversarial attack in the context of images, exploring the adversarial attack in the context of malware is completely different and extremely challenging due to the highly structured nature of software files like PE files.
To put it simply, even though we can employ existing feature-space attacks to generate the ``adversarial features of malware'', it is significantly challenging to find the corresponding ``adversarial malware file'' that can preserve the format, executability, and maliciousness the same as the original malware file.

\textbf{Related work to this survey.}
In fact, for the general adversarial attacks and defenses, there are lots of surveys being done on the image, audio, video, text, and graph~\cite{akhtar2018threat,chakraborty2018adversarial,zhang2019adversarial,serban2020adversarial,machado2021adversarial,long2022survey,alshemali2020improving,lan2022adversarial,zeng2020openattack,sun2018adversarial}, but very few surveys focusing on the adversarial attacks in the context of malware~\cite{pierazzi2020intriguing,park2020survey,li2021arms}.
Here, we introduce those closely related surveys and highlight their limitations and differences compared with our survey paper as follows.
\begin{itemize}
    \item 
    In~\cite{pierazzi2020intriguing}, {Pierazzi~\etal} are the first to present a general mathematical formalization of adversarial attacks in the problem-space and further propose a novel problem-space attack against Android malware detection.
    Although they identify four key constraints and commonalities among different domains (\ie, image classification, face recognition, code attribution, PDF malware, android malware, \etc) in terms of the problem-space adversarial attack, this paper~\cite{pierazzi2020intriguing} is \textbf{not a survey paper} as it neither extensively collects all existing research efforts nor systematically categorizes and summarizes these efforts in this research direction.
    \item
    In~\cite{park2020survey}, {Park and Yener} actually conduct \textbf{an incomplete and improper survey} in reviewing existing adversarial attacks against malware classifiers, since the authors mistakenly categorize them into two categories: gradient-driven and problem-driven approaches, which are clearly not sufficient to cover all existing adversarial attacks against malware detection.
    For instance, the semantics-preserving reinforcement learning (SRL) attack proposed by~\citet{zhang2020semantic} is neither a gradient-driven nor a problem-driven method.
    In addition, this paper~\cite{park2020survey} lacks surveying existing defense methods against such adversarial attacks.
    \item
    In~\cite{li2021arms}, {Li~\etal} first make a series of formulations and assumptions in the context of adversarial malware detection and then attempt to survey this research field of adversarial malware detection in a broad spectrum of malware formats, including Windows PE, Android Package Kit (APK), and Portable Document Format (PDF), which is supposed to be \textbf{too coarse-grained} to fully understand the unique challenges adversarial attacks and defenses for different malware formats.
    It is well-known that malware detection relies heavily on the specific file formats of malware, and thus existing literature papers on malware analysis commonly point out what specific operating system they target and what file format they are~\cite{ucci2019survey,raff2020survey}.    
    Therefore, our paper instead focuses on the malware format of Windows PE, which allows us to specifically identify the distinct characteristics of Windows PE malware and further gain a deeper and thorough understanding of adversarial attacks and defenses in terms of Windows PE malware.
\end{itemize}

\textbf{Contributions of this survey.}
Motivated by the ever-increasing attention of adversarial malware detection, the purpose of this survey is to provide a comprehensive review on the state-of-the-art research efforts of adversarial attacks against Windows PE malware detection as well as corresponding defenses to increase the robustness of existing PE malware detection solutions.
We expect that this survey can serve successive researchers and practitioners who are interested in attacking and defending Windows PE malware detection.
In addition, this survey also aims to provide the basic principles to solve this challenging question in generating ``real'' adversarial PE malware rather than unpractical adversarial PE features that violate the principles (\ie, format-preserving, executability-preserving, and maliciousness-preserving).
We believe that these principles can constitute a useful guideline when related researchers and practitioners deal with the generic task of malware detection, not only restricted to PE malware detection.
To summarize, the key contributions of this survey are as follows.
\begin{itemize}
    \item To the best of our knowledge, this is the first work that summarizes and highlights three unique challenges of adversarial attacks in the context of Windows PE malware in the wild, \ie, format-preserving, executability-preserving, and maliciousness-preserving.
    \item We conduct a comprehensive and systematic review for adversarial attacks against Windows PE malware detection and propose a complete taxonomy to categorize the state-of-the-art adversarial attack methods from different viewpoints.
    \item We summarize the existing adversarial defenses for PE malware detection against the proposed adversarial attacks.
    In addition, we discuss other types of attacks against Windows PE malware detection beyond adversarial attacks.
\end{itemize}

\textbf{Organization.}
The rest of this paper is organized as follows.
We introduce the general layout of PE files and outline the ML/DL-based learning framework of PE malware detection in \Secref{sec:MLDLforPEMalwareDetection}.
In \Secref{sec:challenges}, we manifest three unique challenges of adversarial attacks against PE malware detection compared with the general adversarial attacks in the context of images.
\Secref{sec:adversarial_attacks} first presents our taxonomy of adversarial attacks against PE malware detection and then gives a detailed introduction to the state of the art.
We summarize the existing adversarial defenses for PE malware detection in \Secref{sec:adversarial_defenses}.
In \Secref{sec:discussion}, we first discuss other types of attacks against PE malware detection beyond the adversarial attacks and then point out some research directions and possible opportunities for future work.
We conclude our survey paper in \Secref{sec:conclusion}.
\section{Machine Learning and Deep Learning for PE Malware Detection}\label{sec:MLDLforPEMalwareDetection}

This section aims to provide the basics to understand how to take advantage of ML and DL for malware detection in terms of PE files in the family of Windows operating systems (OSs).
In particular, we first introduce the general layout of PE files and PE malware in \Secref{subsection:PE_file_layout_and_malware}, and then outline the general learning framework of PE malware detection models based on ML/DL in \Secref{subsection:learning_framework_for_PE_malware_detection}.

\subsection{PE file Layout and Malware}\label{subsection:PE_file_layout_and_malware}

\subsubsection{General Layout of PE files}

\begin{wrapfigure}{R}{0.3\textwidth}
    \centering
    \includegraphics[width=0.3\textwidth,keepaspectratio]{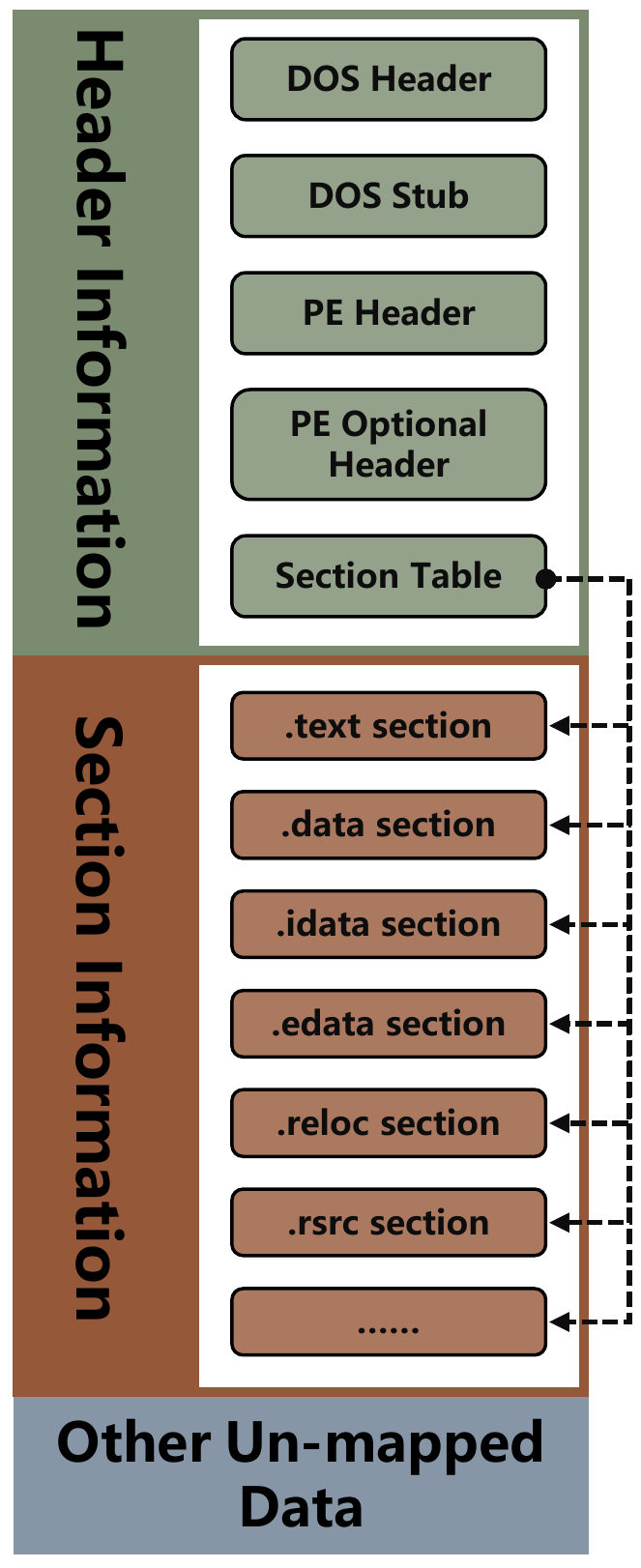}
    \caption{The general layout of a PE file consists of three groups of information: header information, section information, and the other un-mapped data.}
    \label{figure:PE_Format}
\end{wrapfigure}

The Portable Executable (PE) format~\cite{pe_format_2020,pietrek_pe_2020} is created to provide a common file format for the Microsoft family of Windows OSs to execute code and store essential data that is necessary to execute the code on all supported CPUs, which has made PE the dominant file format of executables on the Windows platform since its creation.
As shown in \Figref{figure:PE_Format}, the general file layout of a PE contains three groups of information, \ie, \textbf{header information} (avocado green), \textbf{section information} (sandy brown) and the \textbf{other un-mapped data} (light blue).

The first group is \textbf{header information} which is organized in a fixed manner to tell the OS how to map the PE file into memory for execution.
In particular, every PE file starts with the \textit{DOS Header} followed by the \textit{DOS Stub}, both of which are only used for compatibility with the Microsoft disk operating system (MS-DOS)~\cite{paterson1983inside}.
Following the DOS Header and DOS Stub, the \textit{PE Header} outlines the overview information about the entire PE file, including the number of sections, the creation time, and various other attributes of PE, \etc
Besides that, the subsequent \textit{PE Optional Header} is used to supplementally describe PE files with more than 30 attributes (\eg, size of code, address of entry point, \etc) in more detail.
The last component in the header information group usually is the \textit{Section Table}, which provides information about all associated sections, including names, offsets, sizes, and other information.
It is worth noting that one PE needs at least one section to be loaded and run.

The second group is \textbf{section information} which contains a list of consecutive sections with the executable code and necessary data.
In general, the number and the order of these sections are not fixed.
Although the name of a section can be customized by the user, Microsoft officially defines several naming conventions based on its semantics and functionalities.
In most cases, the ``.text'' section in PE contains the executable code, while the ``.data'' section contains necessary data, mainly storing global variables and static variables.
The ``.idata'' and ``.edata'' sections are used to store the address and size information for the import table and export table, respectively.
To be specific, the import table specifies the APIs that will be imported by this executable, while the export table specifies its own functions so that other executables can access them.
The ``.reloc'' section has relocation information to ensure the executable is positioned and executed correctly.
The ``.rsrc'' section contains all resources (\eg, icons, menus, \etc).
The last group is the \textbf{other un-mapped data} which will not be mapped into memory.
In particular, the un-mapped data refers to the chunk of unused bytes, like debug information, at the end of PE files.

Within the family of Windows OSs, PE mainly has two typical and most commonly used file types, \ie, EXEcutable (EXE) and Dynamic Link Library (DLL), which are generally ended with ``.exe'' and ``.dll'' as the suffix name.
Normally, an ``.exe'' file can be run independently while a ``.dll'' file contains the library of functions that other executables can use in the Windows platform~\cite{pietrek_pe_2020}.

\subsubsection{PE Malware}
Malicious software, \ie, malware, is purposely designed and implemented to satisfy the malicious goals and intentions of attackers, \eg, accessing the system without user permissions, stealing private or confidential information, asking for a large ransom, \etc
Since the PE file format was first created in the family of Windows OSs, PE files have been widely leveraged by malicious attackers to build PE malware.
Until now, according to the security reports from AV-TEST Institute~\cite{avtest_2020_report} and Avira Protection Labs~\cite{avira_2020_q4_report}, PE malware still remains the predominant threat for both personal users and business users in the wild for the following two major reasons.
First, the worldwide popularity of the family of Windows OSs and the commonality of PE files inside make the family of Windows OSs, especially the PE file that can be executed, become the main target of attackers for benefit maximization.
Second, unlike other file types, PE files can be self-contained, which means that PE malware can include all needed data and does not require additional data to launch the attack.
In addition, based on different types of proliferation and different malicious intentions, PE malware can be further briefly classified as viruses, trojans, PUA, worms, adware, ransomware, \etc
For more details, we refer interested readers to~\cite{souppaya2013guide,zeidanloo2010all,ye2017survey}.

\subsection{Learning Framework for PE Malware Detection}\label{subsection:learning_framework_for_PE_malware_detection}
\subsubsection{Overview}
To defend against PE malware, \ie, effectively and efficiently detecting PE malware so that potential infections and damages can be mitigated or stopped, there are tremendous research efforts have been made for PE malware detection.
Traditional malware detection can actually be traced back to classic signature-based malware detection.
To be specific, signature-based malware detection typically maintains a database of signatures of malware that have been previously collected and confirmed.
For a given suspicious software like a PE file, signature-based malware detection can determine whether it is malicious or not by comparing its signature with all signatures from the maintained database of malware.
Obviously, signature-based malware detection has a fatal drawback in that it heavily relies on the maintained database of malware signatures, so that it can only detect previously collected and known malware.

\begin{figure*}
    \centering
    \includegraphics[width=0.9\textwidth,keepaspectratio]{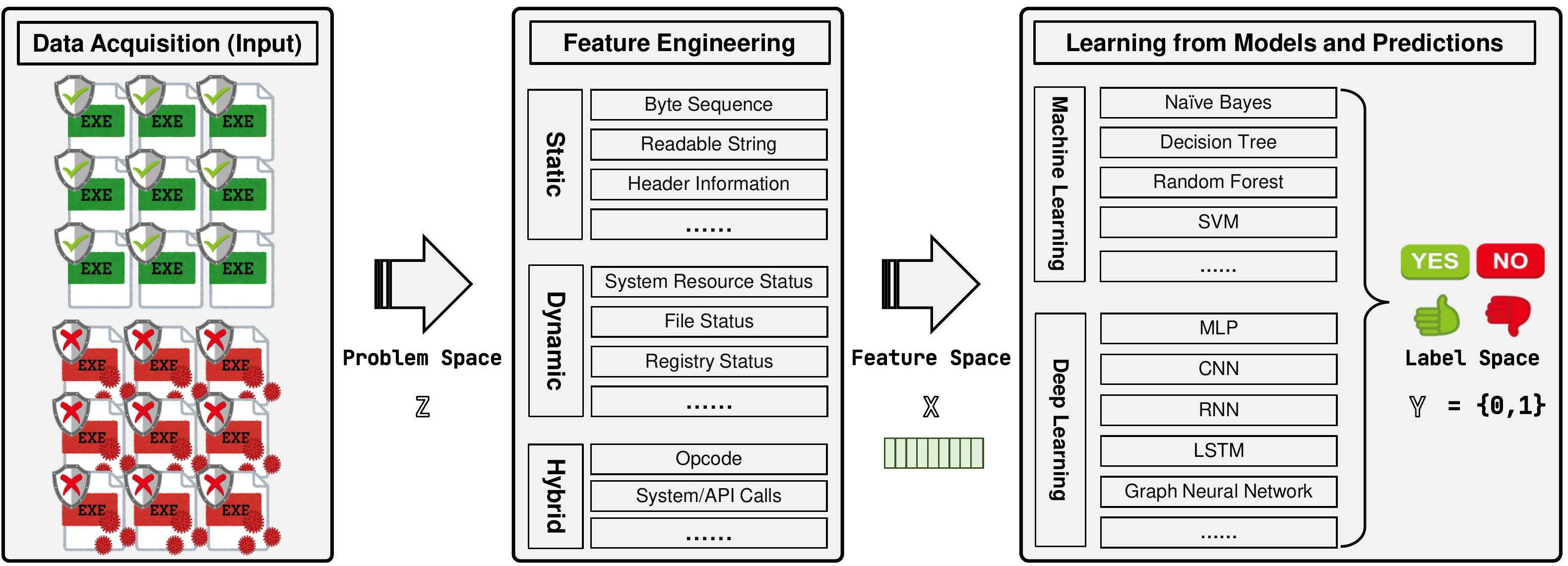}
    \caption{The Overview Learning Framework of PE Malware Detection.}
    \label{figure:malware_detection_model_overview}
    \vspace{-3mm}
\end{figure*}

In recent years, inspired by the great successes of ML and DL techniques in various research fields,
various ML/DL-based malware detection methods that leverage the high learning capacity of ML/DL models have been adapted and proposed for PE malware detection.
These ML/DL-based malware detection methods normally claim that, as ML/DL models generalize well to predictions of new and unseen instances, they can also generalize to new and previously unseen (\ie, zero-day) malware.
\Figref{figure:malware_detection_model_overview} illustrates the overview learning framework of PE malware detection~\cite{ceschin2020machine}, which generally consists of three steps, including data acquisition, feature engineering as well as learning from models and predictions.
In the following, we are going to introduce each step at a glance.

\subsubsection{Data Acquisition}
It is well known that the data quality determines the upper limit of ML/DL models~\cite{cortes1995limits}.
In order to build any malware detection models, it is fundamental to collect and label sufficient PE samples, including both malware and goodware of PE files.
However, unlike the benchmark datasets in the field of computer vision or natural language processing (\eg, MNIST~\cite{lecun1998gradient_mnist}, CIFAR~\cite{krizhevsky2009learning_cifar}, ImageNet~\cite{deng2009imagenet}, Yelp Review~\cite{yelp_dataset}, \etc), cybersecurity companies or individuals normally treat PE samples, especially raw files of PE malware, as private property, and thus rarely release them to public.
Although a few security institutions or individuals have shared their datasets of PE files~\cite{virusshare,thezoo}, there is no established benchmark that has been widely used for most PE malware detection until now.
In addition to collecting PE samples, another important process is to distinguish PE malware from all collected PE files, \ie, data labeling.
Currently, a common practice to label PE files is to rely on malware analysis services like VirusTotal, which can provide multiple detection results (\ie, whether it is malware) from nearly 70 state-of-the-art anti-virus engines~\cite{chen2015finding}.
Unfortunately, for the same suspicious PE sample, it is well-known that its detection results from different anti-virus engines are somewhat inconsistent with each other.
To address these challenges, a variety of methods~\cite{sebastian2016avclass,sebastian2020avclass2,kantchelian2015better,zhu2020benchmarking} have been proposed to unify the detection label by either picking up a ``credible'' anti-virus (\ie, the most effective and recognized anti-virus software like Kaspersky or Norton) or adopting a voting-based approach from multiple anti-virus engines.

Formally, suppose there is a \textbf{problem space} $\mathbb{Z}$ (also known as input space) that contains objects of one specific domain (\eg, PE files, images, texts, \etc) from the real-world applications, each object $z$ in the problem space $\mathbb{Z}$ ($z \in \mathbb{Z}$) is associated with a ground-truth label $y \in \mathbb{Y}$, in which $\mathbb{Y}$ denotes the corresponding \textbf{label space} depending on the specific real-world application.
In the scenario of PE malware detection, the problem space $\mathbb{Z}$ is referred to the space of all possible PE files, and the label space $\mathbb{Y}$ (\eg, $\mathbb{Y} = \{-1, 1\}$ or $\mathbb{Y} = \{0, 1\}$) denotes the space of detection labels (\eg, $-1 \text{ or } 0$ denotes malware and $1$ denotes goodware.)

\begin{table*}[htbp]
  \centering
  \small
  \caption{Common Features and Corresponding Learning Models for PE Malware Detection.}
    \begin{tabular}{>{\centering\arraybackslash}m{28mm}m{5mm}p{5mm}>{\raggedright\arraybackslash} p{46mm}>{\raggedright\arraybackslash}p{48mm}}
    \toprule
    \multirow{2}[12]{*}{Features} & \multicolumn{2}{c}{Category} & \multicolumn{1}{c}{\multirow{2}[12]{*}{\tabincell{c}{Representative ML/DL\\Model Architectures}}} & \multicolumn{1}{c}{\multirow{2}[12]{*}{\tabincell{c}{Representative PE Malware\\Detection Methods$^*$}}} \\
    \cmidrule{2-3} & \multicolumn{1}{c}{\vleftthead{Static}} & \multicolumn{1}{c}{\vleftthead{Dynamic}} &   & \\
    \midrule
    \rowcolor{lightgrayfortable}
      Byte Sequence            & \checkmark &            & Na\"ive Bayes~\cite{murphy2006naive}, SVM~\cite{cortes1995support}, DT~\cite{witten2002data}, MLP~\cite{collobert2004links}, CNN~\cite{krizhevsky2012imagenet}, LightGBM~\cite{ke2017lightgbm}. & \citet{schultz2001data}, \citet{kolter2006learning}, \citet{saxe2015deep},  \citet{gibert2018classification}, MalConv~\cite{raff2017malware}, EMBER~\cite{anderson2018ember}.\\
    
    % \midrule
      Readable String           & \checkmark &            & Na\"ive Bayes~\cite{murphy2006naive}, SVM~\cite{cortes1995support}, MLP~\cite{collobert2004links}, DT~\cite{witten2002data}, RF~\cite{ho1995random}, LightGBM~\cite{ke2017lightgbm}. & \citet{schultz2001data}, SBMDS~\cite{ye2009sbmds}, \citet{islam2010classification}, \citet{saxe2015deep}, EMBER~\cite{anderson2018ember}.\\
    
    % \midrule
      \rowcolor{lightgrayfortable}
      Header Information        & \checkmark &            & MLP~\cite{collobert2004links}, LightGBM~\cite{ke2017lightgbm}, Na\"ive Bayes~\cite{murphy2006naive}, SVM~\cite{cortes1995support}, DT~\cite{witten2002data}. & \citet{saxe2015deep}, EMBER~\cite{anderson2018ember}, PE-Miner~\cite{shafiq2009pe_miner}.\\
    
    % \midrule
      Grayscale Image           & \checkmark &            & kNN~\cite{altman1992introduction_knn}, CNN~\cite{krizhevsky2012imagenet}, SVM~\cite{cortes1995support}. & \citet{nataraj2011comparative}, \citet{kim2017image}, Visual-AT~\cite{liu2020novel}.\\
    
    % \midrule
    \rowcolor{lightgrayfortable}
      CPU/Memory/IO {\textit{etc.} Status} &            & \checkmark & CNN~\cite{krizhevsky2012imagenet}, kNN~\cite{altman1992introduction_knn}, SVM~\cite{cortes1995support}. & \citet{rieck2008learning}, AMAL~\cite{mohaisen2015amal}, \citet{abdelsalam2018malware}. \\
    
    % \midrule
      File Status               &            & \checkmark & Hierarchical Clustering~\cite{franklin2005elements}, CNN~\cite{krizhevsky2012imagenet}, kNN~\cite{altman1992introduction_knn}, SVM~\cite{cortes1995support}, GNN~\cite{kipf2016semi}.  & \citet{bailey2007automated}, \citet{rieck2008learning}, AMAL~\cite{mohaisen2015amal}, \citet{abdelsalam2018malware}, MatchGNet~\cite{Heterogeneous2019}. \\              
    
    % \midrule
    \rowcolor{lightgrayfortable}
      Registry Status           &            &  \checkmark & Hierarchical Clustering~\cite{franklin2005elements}, kNN~\cite{altman1992introduction_knn}, SVM~\cite{cortes1995support}. & \citet{bailey2007automated}, \citet{rieck2008learning}, AMAL~\cite{mohaisen2015amal}. \\
      
    % \midrule
      Network Status            &            & \checkmark & Hierarchical Clustering~\cite{franklin2005elements}, CNN~\cite{krizhevsky2012imagenet}, kNN~\cite{altman1992introduction_knn}, SVM~\cite{cortes1995support}, GNN~\cite{kipf2016semi}. & \citet{bailey2007automated}, \citet{rieck2008learning}, AMAL~\cite{mohaisen2015amal}, \citet{abdelsalam2018malware}, MatchGNet~\cite{Heterogeneous2019}. \\

    % \midrule
    \rowcolor{lightgrayfortable}
      Opcode                    & \checkmark & \checkmark & kNN~\cite{altman1992introduction_knn}, SVM~\cite{cortes1995support}, DT~\cite{witten2002data}, RNN~\cite{cho2014learning_rnn}, CNN~\cite{krizhevsky2012imagenet}, Hierarchical Clustering~\cite{franklin2005elements}. &  AMCS~\cite{ye2010automatic}, \citet{santos2013opcode}, IRMD~\cite{zhang2016irmd}, RMVC~\cite{sun2018deep}. \\
    
    % \midrule
      System or API Calls       & \checkmark & \checkmark & RIPPER~\cite{cohen1996learning_ripper}, SVM~\cite{cortes1995support}, Hierarchical Clustering~\cite{franklin2005elements}, CNN~\cite{krizhevsky2012imagenet}, LSTM~\cite{lstm:hochreiter1997long}. & \citet{schultz2001data}, SBMDS~\cite{ye2009sbmds}, \citet{rieck2008learning}, \citet{qiao2013analyzing}, \citet{zhang2020dynamic}. \\
    
    % \midrule
    \rowcolor{lightgrayfortable}
      Control Flow Graph        & \checkmark & \checkmark &  Na\"ive Bayes~\cite{murphy2006naive}, SVM~\cite{cortes1995support}, RF~\cite{ho1995random}, GNN~\cite{kipf2016semi}.  & \citet{kapoor2016control}, MAGIC~\cite{yan2019classifying}, MalGraph~\cite{ling2022malgraph}. \\
    
    % \midrule  
      Function Call Graph       & \checkmark & \checkmark & RF~\cite{ho1995random}, AutoEncoder~\cite{vincent2010stacked_autoencoder}, CNN~\cite{krizhevsky2012imagenet}, GNN~\cite{kipf2016semi}. & \citet{hassen2017scalable}, DLGraph~\cite{jiang2018dlgraph}, DeepCG~\cite{zhao2019deepcg}, MalGraph~\cite{ling2022malgraph}. \\
    
    \bottomrule
    \end{tabular}%
    % \\ \flushleft \footnotesize{\noindent $^*$If the paper clearly names the PE malware detection method, we use the model name with its citation, otherwise we use the author name(s) of the paper with its citation.}
    \\  \scriptsize{\noindent $^*$If the paper does not clearly name the PE malware detection, we use the author name(s) of the paper with its reference.}%
  \label{table:malware_feature_model}%
\end{table*}%

\subsubsection{Feature Engineering}\label{subsubsection:feature_engineering}
After collecting and labeling sufficient PE samples, it is necessary and important to perform somewhat feature engineering over all PE samples before inputting them into ML/DL models, as ML/DL models can only accept numeric input.
Feature engineering aims to extract the intrinsic properties of PE files that are most likely to be used for distinguishing malware from goodware, and then generates corresponding numeric features for representation.
From different perspectives of properties of PE files, there is a large body of work on extracting various features, which can be generally categorized into three broad category: \textit{static} features, \textit{dynamic} features and \textit{hybrid} features~\cite{ye2017survey,raff2020survey,ceschin2020machine} and summarized in \Tabref{table:malware_feature_model}.

First of all, static features are directly extracted from the PE samples themselves without actually running them.
For instance, byte sequence, readable string, header information, and the grayscale image are commonly used static features for PE malware detection.

\begin{itemize}
    \item Byte Sequence:
    A PE sample is essentially a binary file, which is typically considered to be a sequence of bytes.
    Therefore, the byte sequence is the most straightforward and informative way to represent a PE file.
    In fact, the byte sequence can either be directly input into DL models~\cite{raff2017malware,krvcal2018deep,coull2019activation}, or be further converted into an intermediate representation, \eg, n-grams or entropy of byte sequences~\cite{schultz2001data,kolter2006learning,saxe2015deep,gibert2018classification}.
    
    \item Readable String:
    A PE file might contain readable strings that reflect it intentions or semantics, like file names, IP addresses, domain names, author signatures, \etc
    After extracting readable strings in a PE file, their numeric feature representation can be a set of binary attributes (\ie, whether the string exists), frequencies, or even 2D histogram features~\cite{schultz2001data,ye2009sbmds,islam2010classification,saxe2015deep}.
    
    \item Header Information:
    As illustrated in \Figref{figure:PE_Format}, the PE header information occupies an important place to describe and normalize the PE file globally so that it can be executed properly.
    In particular, simple statistics on PE header information, such as the file size, numbers of sections, sizes of sections, the number of imported or exported functions, \etc, are commonly used feature representations for PE malware detection~\cite{saxe2015deep,anderson2018ember,shafiq2009pe_miner}.
    
    \item Grayscale Image:
    Since the value range of bytes in a PE file is the same as the pixel value in an image, a visualization-based feature engineering approach is to transform a PE file into a grayscale image, for which each byte in a PE file corresponds to a pixel in an image~\cite{nataraj2011malware}.
    Inspired by the recent great successes of image classification methods, a lot of visualization-based methods have also been proposed for PE malware detection~\cite{nataraj2011comparative,kim2017image,liu2020novel}.
    
\end{itemize}

Second, dynamic features refer to those features that can be extracted by first running the executable in an isolated environment (\eg, sandbox, virtual machine, \etc) and then monitoring their runtime status in terms of system resources, files, registries, network, and others.

\begin{itemize}
    \item System Resource Status:
    The execution of malware inevitably occupies system resources (\eg, CPU, memory, IO, \etc), whose runtime status can be considered as dynamic features for malware detection, as a variety of malware within one specific family might follow a relatively fixed pattern of system resources during execution.
    In particular, CPU usage, memory usage, and I/O request packets are commonly monitored as dynamic features~\cite{rieck2008learning,mohaisen2015amal,abdelsalam2018malware}.
    
    \item File Status:
    Malware normally needs to operate on files of target users for reaching malicious intentions by attackers.
    Thus, logging and counting for the files accessed, created, modified, or deleted are commonly used dynamic features in malware detection~\cite{bailey2007automated,rieck2008learning,mohaisen2015amal,abdelsalam2018malware,Heterogeneous2019}.
    
    \item Registry Status:
    Registries that store the system/application-level configurations are important for the family of Windows OSs.
    The malware could operate on registries with malicious intentions, like self-starting malware.
    Similar to file status, registry status like counting the registries created, modified, and deleted can also be regarded as dynamic features~\cite{bailey2007automated,rieck2008learning,mohaisen2015amal}.
    
    \item Network Status:
    The spread of malware like trojans and ransomware mainly depends on the network.
    Taking trojans as an example, they are likely to connect remote servers with certain network ports.
    Therefore, when diving into the specific aspects of network status, there is a variety of network-level information that can be used for creating a rich set of dynamic features~\cite{bailey2007automated,rieck2008learning,mohaisen2015amal,abdelsalam2018malware,Heterogeneous2019}, such as the number of distinct IP addresses or certain ports, the number of different HTTP requests (\eg, POST, GET, HEAD, PUT, \etc), the number of common DNS record types (\eg, PTR, CNAMN, SOA, \etc), to name just a few.
    
\end{itemize}

Finally, we exemplify four commonly used hybrid features, \ie, opcode, system/API calls, control flow graph (CFG), and function call graph, which can be extracted from executables with either static analysis methods or dynamic analysis methods.
For instance, opcodes of executables can be obtained by either extracting from their disassembled instructions or monitoring their runtime instructions in memory.

\begin{itemize}
    \item Opcode:
    Executables, including malware, can be generally considered as a collection of instructions that are executed in a specific order.
    In machine assembly language, an instruction consists of an opcode and several operands, in which the opcode specifies the operation to be executed and the operand refers to the corresponding data or its memory location.
    As prior studies suggest, the opcode distributions of malware statistically differ from goodware, and thus various features are constructed from the opcodes, such as their frequency, n-grams of opcode sequences, or even opcode images~\cite{ye2010automatic,santos2013opcode,zhang2016irmd,sun2018deep}.
    
    \item System/API Calls:
    System/API calls refer to how executables interact with system-level or application-level libraries in the family of Windows OSs.
    Similar to the opcode, various feature representations are thus constructed from system/API calls, such as the frequency of system/API calls, and n-grams of system/API call sequences~\cite{schultz2001data,ye2009sbmds,rieck2008learning,qiao2013analyzing,zhang2020dynamic}.
    
    \item CFG:
    The CFG is a graph-based feature representation that is commonly used to characterize the control flow of executables, including PE malware~\cite{kapoor2016control,yan2019classifying}.
    Building from assembly instructions of executables, each node in the CFG represents a sequence of instructions without branching and each edge represents the control flow path between two nodes~\cite{ling2021multilevel}.
    
    \item Function Call Graph:
    The function call graph~\cite{ryder1979constructing} that attempts to build the caller-callee relation between different functions (including system/API or user-implemented functions), is regarded as a more coarse-grained graph representation compared with CFG~\cite{hassen2017scalable,jiang2018dlgraph,zhao2019deepcg}.

\end{itemize}

It is worth noting that, on the one hand, we just briefly review and categorize the commonly used features in PE malware detection, and do not attempt to cover all, which is not the goal of our paper.
On the other hand, all the features mentioned above are not separate or independent, they are actually mixed for PE malware detection in the wild.
In essence, the process of feature engineering can be broadly expressed as a feature mapping function that maps the problem space into the feature space (\ie, the numeric features), which is formulated in \Defref{definition:feature_mapping_function} as follows.

\begin{definition}[Feature Mapping Function]
A feature mapping function $\phi$ is formulated as $\phi: \mathbb{Z} \rightarrow \mathbb{X}$, in which $\mathbb{Z}$ denotes the problem space from a specific real-world application, and $\mathbb{X}$ denotes the corresponding feature space, numerically describing the intrinsic properties of objects in the problem space.
\label{definition:feature_mapping_function}
\end{definition}

\subsubsection{Learning from Models and Predictions}
After extracting and generating the numeric features from the executable, it is necessary to choose a proper ML or DL model for PE malware detection that is generally regarded as a binary classification task, \ie, predicting whether the given executable is malware or goodware.
In recent years, with the rapid development and great success of artificial intelligence technology in many fields like computer vision, natural language processing, and even code analysis~\cite{ling2021deep}, a huge variety of ML/DL models have been continuously proposed, such as Na\"ive Bayes, SVM, DT, RF, MLP, CNN, RNN, LSTM, or GNN.
Regardless of how diverse of feature representation for executables, almost all kinds of ML/DL models mentioned have been used for PE malware detection, as long as the features obtained from feature engineering conform to the input format of the corresponding ML/DL model.
For instance, an LSTM model can accept the sequence data as the input, but cannot accept the graph data, while GNN can process the graph data.

In essence, an ML/DL model refers to a mathematical discrimination function $f$ with parameters to map the numeric features of executables into their binary labels (\ie, malware and goodware), which is broadly formulated in \Defref{equation:discrimination_function} as follows.

\begin{definition}[Discrimination Function]
A discrimination function $f$ can be precisely formulated as $f: \mathbb{X} \rightarrow \mathbb{Y}$, in which $\mathbb{X}$ denotes the feature space and $\mathbb{Y}$ denotes the corresponding label space.
\label{equation:discrimination_function}
\end{definition}

The training process of a malware detection model is to learn the model parameters based on a large number of training samples, so that the malware detection model can approximate the real relationship function between the feature patterns of executables and their binary detection labels.
After that, to predict whether a given executable is malware or not, the malware detection model with learned parameters can effectively and efficiently compute the probabilities assigned to both classes of malware and goodware.
In order to find the most applicable model, it is actually quite common to test different ML/DL models for PE malware detection depending on the specific task.
In \Tabref{table:malware_feature_model}, the last two columns present the representative ML/DL model architectures and corresponding PE malware detection methods with references.
\section{Challenges of Adversarial Attacks for PE Malware}\label{sec:challenges}

In this section, we first introduce the general concept and taxonomy of adversarial attacks that have been originally and extensively studied in the domain of image classification tasks, and then manifest the most unique challenges of adversarial attacks for PE malware when compared with other fields like images, audios, texts, \etc

\subsection{Adversarial Attacks: The General Concept and Taxonomy}\label{subsec:adversary_attack_concept_taxonomy}
Although recent advances in ML and DL have led to breakthroughs in a variety of long-standing real-world tasks (\eg, computer vision, natural language processing, speech recognition, \etc), unfortunately, it has been convincingly demonstrated that existing ML and DL models are inherently vulnerable to adversarial attacks with carefully crafted adversarial examples.
In particular, adversarial examples are intentionally and maliciously designed inputs that aim to mislead the given target model in the testing phrase rather than the training phrase.
Generally, adversarial attacks can be categorized along multiple different dimensions.
In the following part, we broadly classify adversarial attacks along two dimensions, \ie, adversary's space and adversary's knowledge.

\begin{figure*}[htb]
    \centering
    \includegraphics[width=0.9\textwidth,keepaspectratio]{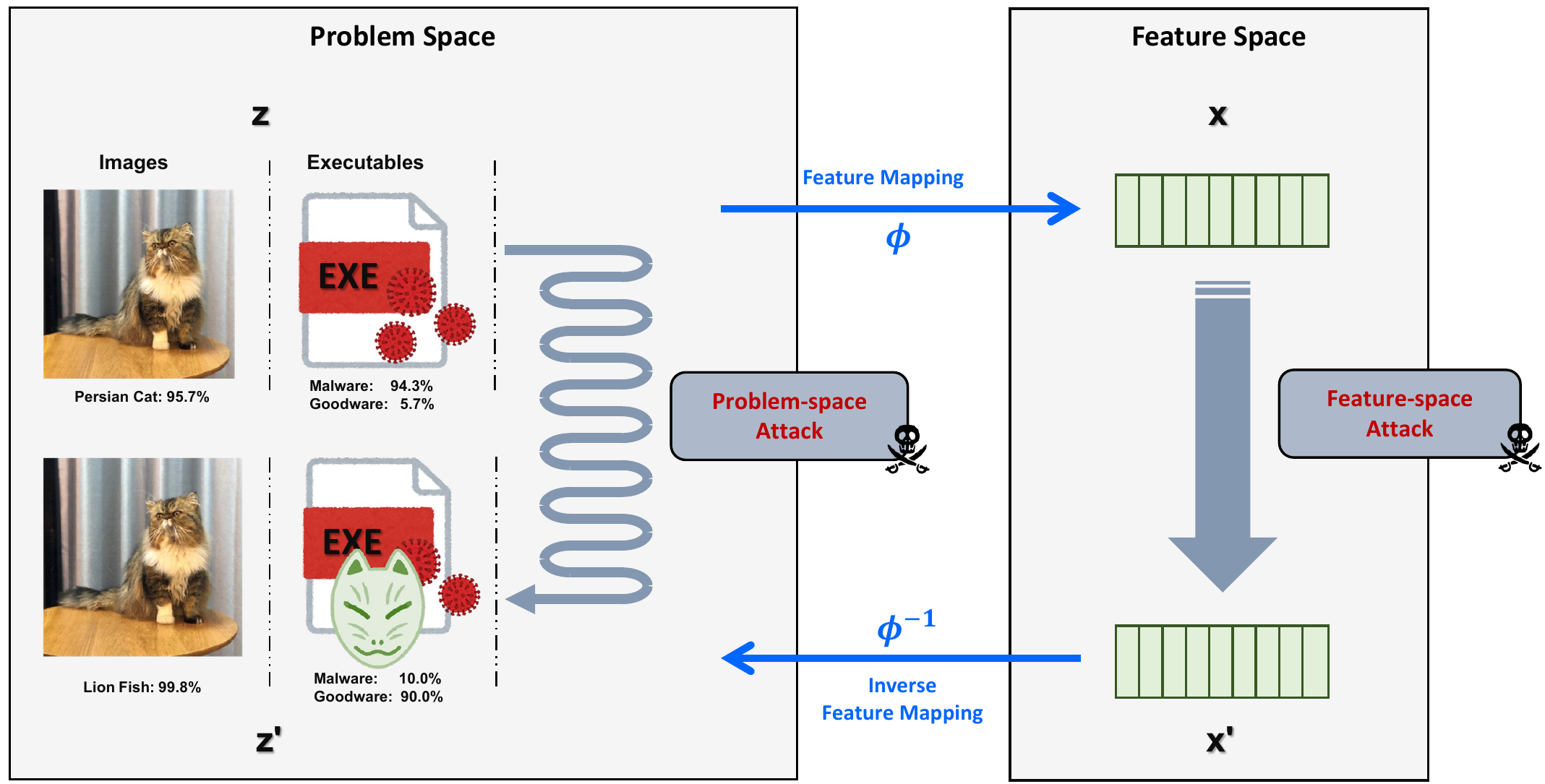}
    \caption{Illustration of the connection between the \textbf{feature-space attack} and the \textbf{problem-space attack}, in which the feature mapping function $\phi$ and the inverse feature mapping function $\phi^{-1}$ act as bridges for transitions between the feature-space and the problem-space.}
    \label{figure:problem_feature_connection}
\end{figure*}

\subsubsection{Adversary's Space: Feature-space Attack versus Problem-space Attack}
Initially, the adversarial example is explored in the context of image classification tasks and is normally addressed in the feature-space domain.
In particular, for a given image classification model $f$ and a given input image $z$ with feature representation $x \in \mathbb{X}$ (\ie, $f(x) = y$), the attacker attempts to minimize the distance between $x^{\prime}$ and $x$ in the feature-space, such that the resulting adversarial example $x^{\prime} \in \mathbb{X}$ in the feature-space can misclassify the classification model $f$.
This kind of adversarial attack is normally termed as the \textbf{feature-space attack}, which is formulated in \Eqref{equation:adversarial_example_feature} as follows.
\begin{equation}
\centering
\begin{aligned}
    \min_{x^{\prime}}    & \  \text{distance}(x^{\prime}, x)  \\
    \text{s.t.}          & \  x^{\prime} \in \mathbb{X}       \\
                         & \  f(x^{\prime}) = y^{\prime} \neq y
\end{aligned}
\label{equation:adversarial_example_feature}
\end{equation}
in which $\text{distance}(x_1, x_2)$ denotes any measurement function of distance between $x_1$ and $x_2$ in the feature-space depending on the actual applications.
For the feature-space attacks, as the feature representation in the feature-space is normally continuous, most of them generate adversarial examples based on gradients-base methods like FGSM, PGD, C\&W, \etc

In contrast with the aforementioned feature-space attack, the \textbf{problem-space attack} refers to the adversarial attack that is performed in the problem-space, \ie, how to ``modify'' the real-world input $z \in \mathbb{Z}$ with a minimal cost (\eg, executables, source code, PDF,\etc) such that the generated adversarial example $z^{\prime} \in \mathbb{Z}$ in the problem-space can also misclassify the target model $f$ as follows.
\begin{equation}
\centering
\begin{aligned}
    \min_{z^{\prime}} & \ \text{cost}(z^{\prime}, z) \\
    \text{s.t.}  & \ z^{\prime} \in \mathbb{Z} \\
                 & \ f(\phi(z^{\prime})) = y^{\prime} \neq y
\end{aligned}
\label{equation:adversarial_example_problem}
\end{equation}
in which $\text{cost}(z_2, z_1)$ denotes any cost function that transforms $z_1$ into $z_2$ in the problem-space of the specific application.

When comparing the problem-space attack in \Eqref{equation:adversarial_example_problem} with the feature-space attack in \Eqref{equation:adversarial_example_feature}, 
it is easy to find the most fundamental and noticeable difference between them is, the problem-space attack involves a feature mapping function $\phi$ that maps the problem-space into the feature-space, which is usually neither invertible nor differentiable.
Therefore, the problem-space attack can hardly use gradient-based methods directly to generate adversarial examples.
In \Figref{figure:problem_feature_connection}, we illustrate the connection between the feature-space attack and the problem-space attack.

\subsubsection{Adversary's Knowledge: White-box Attack versus Black-box Attack}
Adversary's knowledge specifies what we assume the adversary knows about the target model to be attacked.
In terms of adversary's knowledge, adversarial attacks can be further categorized into the \textbf{white-box attack} and the \textbf{black-box attack}.
To be specific, the white-box attack refers to the scenario that the attackers know all information about the target model (\eg, architectures, weights/parameters, outputs, features, \etc) as well as the dataset to train the target model.
By contrast, the black-box attack refers to the scenario that the attackers know nothing about the target model except the model output, \eg, the classification label with or without probability.\footnote{There is no unified view on whether to treat the scenario of knowing the classification label with probability as the black-box attack.}
Apparently, the black-box attack is much harder to satisfy than the white-box attack since the white-box attack is equipped with more knowledge about the target model.
Besides, there is a wide spectrum between the white-box attack and the black-box attack, which is usually broadly referred as the \textbf{gray-box} attack.
Therefore, for any gray-box attack, it is significantly necessary to show to what extent the adversary knows and does not know about the target model as well as the training dataset.

\subsection{Three Unique Challenges of Adversarial Attacks for PE Malware: From Feature-space to Problem-space}
Originally, adversarial attacks are explored in the domain of image classification tasks and a variety of feature-space attacks are subsequently proposed to generate adversarial examples for the malicious purpose of misclassification, \eg, misclassifying a Persian cat into a lionfish with a high probability of 99.8\% as depicted in \Figref{figure:problem_feature_connection}.
Actually, the main reason for the success of directly performing the feature-space attack to generate adversarial examples of images is that, it is easy to find the corresponding image $z^{\prime}$ from the generate adversarial feature $x^{\prime}$ via the inverse feature mapping function $\phi^{-1}$ (\ie, $z^{\prime} = \phi^{-1} (x^{\prime})$), as indicated in \Figref{figure:problem_feature_connection}.
However, when considering the adversarial attacks for the PE files, the circumstance becomes completely different and extremely challenging due to the \textbf{problem-feature space dilemma}~\cite{quiring2019misleading}, which is mainly manifested in the following two aspects.

\begin{enumerate}
    \item The feature mapping function $\phi_{image}$ for images is relatively fixed (\ie, an image can be formatted as a two-dimensional array of pixels where each pixel value is a three-dimensional RGB vector with a continuous value between 0 to 255), while the feature mapping function $\phi_{pe}$ for PE files is not fixed and can take various and diverse approaches of feature engineering as detailed in \Secref{subsubsection:feature_engineering}.
    Especially in the setting of black-box attacks, the attacker cannot know the specific feature mapping function $\phi_{pe}$ for PE files, which greatly increases the difficulty of adversarial attacks for PE files.
    
    \item For images, although the inverse feature mapping function $\phi^{-1}_{image}$ is not exactly bi-injective (\eg, the pixel value might not be in the range of 0 to 255), it is continuously differentiable, and thus the feature-space attack based on gradients can directly apply on images to generate adversarial examples.
    However, for various different feature mapping functions of PE files, to map a feature vector in the feature-space into an executable in the problem-space, it is almost impossible to find an exact or approximate function of inverse feature mapping $\phi^{-1}_{pe}$ that is either bi-injective or differentiable.
\end{enumerate}

\begin{figure}[ht]
    \begin{minipage}[ht]{0.55\textwidth}
        \centering
        \includegraphics[width=0.99\textwidth]{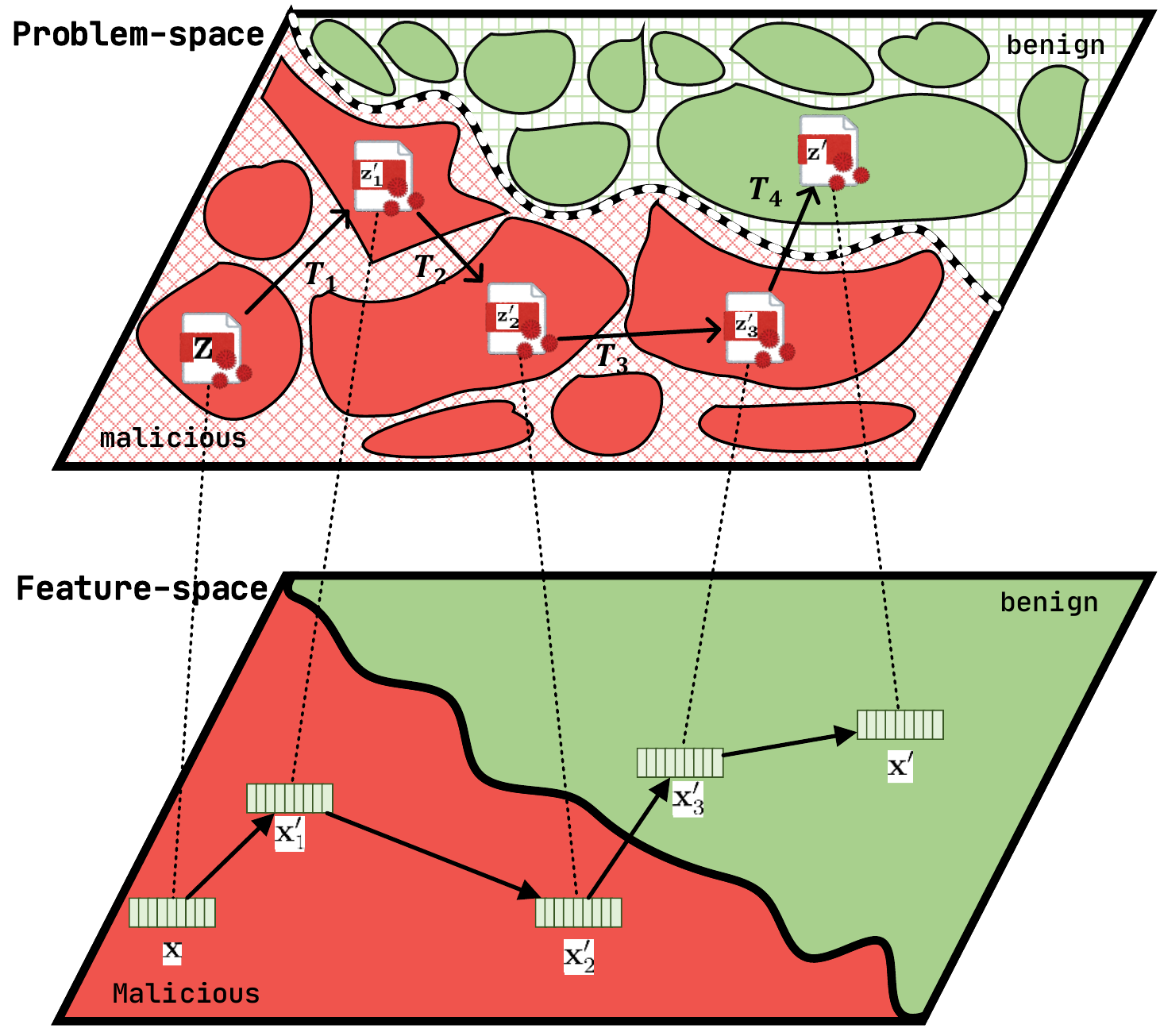}
        \caption{The schematic illustration of the \textbf{feature-space attack} versus the \textbf{problem-space attack} for PE malware, in which the original PE malware $z$ is manipulated in the problem-space to continuously generate the adversarial PE malware (\ie, $z^{\prime}_{1}$, $z^{\prime}_{2}$, $z^{\prime}_{3}$ and $z^{\prime}$), while the corresponding PE malware feature $x$ in the feature-space is mapped to continuously generate adversarial PE malware features (\ie, $x^{\prime}_{1}$, $x^{\prime}_{2}$, $x^{\prime}_{3}$ and $x^{\prime}$).}
        \label{figure:problem_feature_attack}
    \end{minipage}
    % \hfill
    \hspace{3mm}
    \begin{minipage}[ht]{0.42\textwidth}
        \centering
        \includegraphics[width=0.99\textwidth]{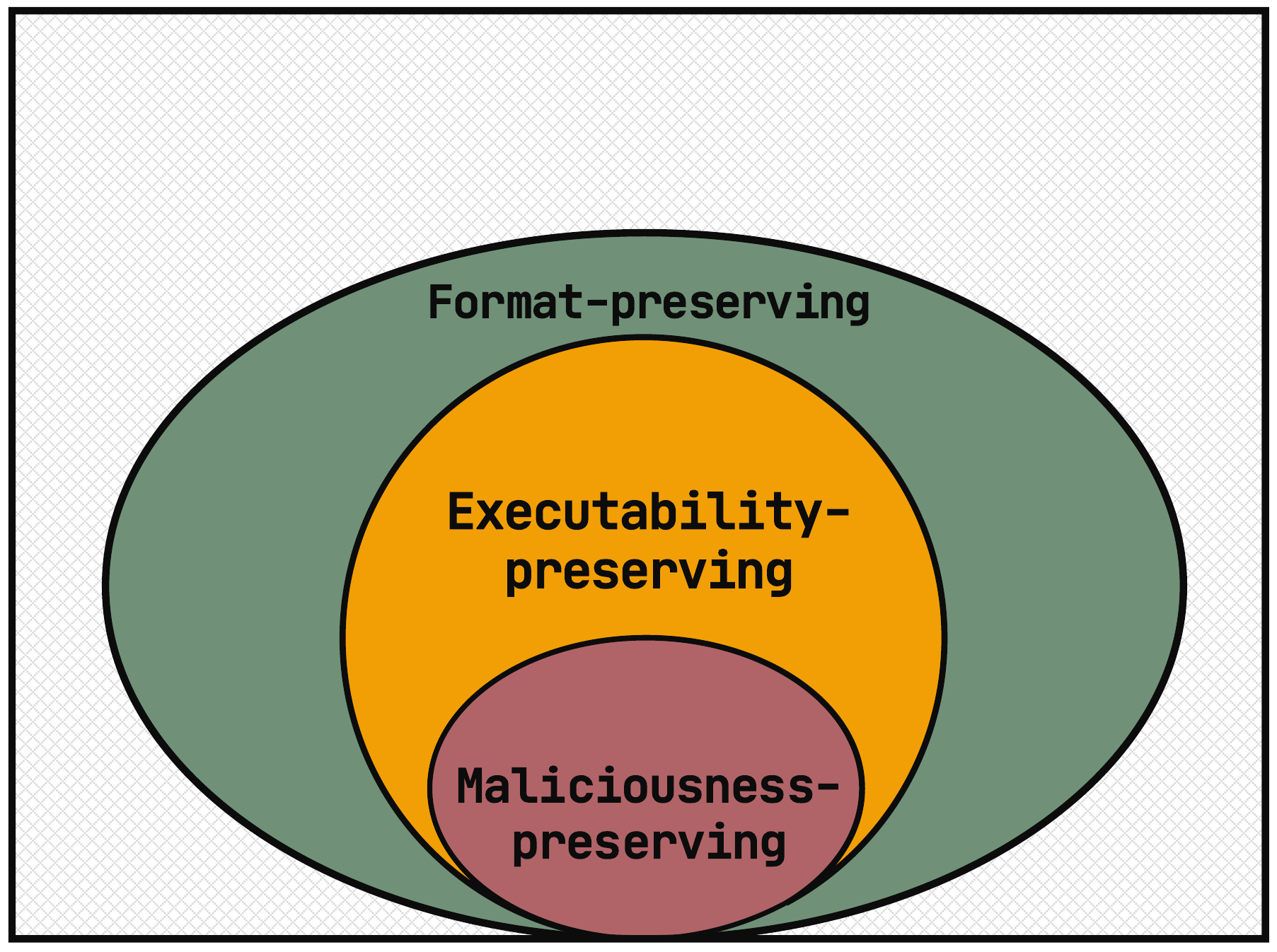}
        \caption{The schematic illustration of relationship between the three unique challenges of adversarial attacks for PE malware, \ie, \textbf{format-preserving}, \textbf{executability-preserving} and \textbf{maliciousness-preserving}.}
        \label{figure:Challenges}
    \end{minipage}
\end{figure}

As depicted in \Figref{figure:problem_feature_attack}, in order to generate adversarial examples for PE files, although there is a variety of adversarial attacks that exploit the feature-space attacks based on gradients have been proposed, we argue that these adversarial attacks are \textit{impractical} and \textit{unrealistic} against PE malware detection in the wild world.
This is because what these adversarial attacks generate is the ``adversarial PE feature'' rather than the ``adversarial PE malware'' in the end, and an ``adversarial PE feature'' does not guarantee to correspond to an ``adversarial PE malware'' due to the following two reasons.
On the one hand, it is almost impossible to find a corresponding adversarial PE malware $z^{\prime}$ based on the generated adversarial PE feature $x^{\prime}$, as the inverse feature mapping function $\phi^{-1}_{pe}$ is normally neither bi-injective nor differentiable.
On the other hand, even though we could find the exact PE malware $z^{\prime}$ in the problem-space that corresponds to the generate adversarial feature $x^{\prime}$ in the feature-space, there is no guarantee that the found $z^{\prime}$ is also ``adversarial''.
Taking the $x^{\prime}_3$ in \Figref{figure:problem_feature_attack} as an example, although its feature representation $x^{\prime}_3$ in feature-space is misclassified as benign (\ie, $f(x^{\prime}_3) = 0$), but its corresponding PE malware object $z^{\prime}_3$ in problem-space is still detected as malicious (\ie, $f(\phi(z^{\prime}_3)) = 1 \neq 0$).

Therefore, to further generate practical and realistic adversarial PE malware against malware detection in the wild, one of the possible or even the only way so far is to seek for the problem-space attack to generate adversarial PE malware in the problem-space as defined in \Eqref{equation:adversarial_example_problem}.
To be specific, as depicted in \Figref{figure:problem_feature_attack}, current problem-space attacks normally attempt to find and apply a series of self-defined problem-space transformations (\ie, $\mathbf{T}_1$, $\mathbf{T}_2$, $\mathbf{T}_3$ and $\mathbf{T}_4$) that sequentially transform the original PE malware $z$ into the desired adversarial PE malware $z^{\prime}$ (\ie, $z \xrightarrow[]{\mathbf{T}_1} z^{\prime}_{1} \xrightarrow[]{\mathbf{T}_2} z^{\prime}_{2} \xrightarrow[]{\mathbf{T}_3} z^{\prime}_{3} \xrightarrow[]{\mathbf{T}_4} z^{\prime}$), such that
\ding{182}~$z^{\prime}$ is no longer detected as malicious by the target malware detection,
and
\ding{183}~$z^{\prime}$ maintains the same semantics as the original $z$.
In the following parts, we detail the three unique challenges of maintaining the semantics of adversarial PE malware for practical and realistic adversarial attacks against PE malware detection and present the relationship between the three challenges in \Figref{figure:Challenges}.

\subsubsection{Challenge 1: Follow the format of PE files (\textbf{format-preserving})}
First of all, unlike images, audio, or even texts, PE malware must follow the standard and strict format rules of PE files.
As characterized in \Figref{figure:PE_Format}, the PE file normally has a relatively fixed layout and structure that is necessary to first load the PE file in the system memory and then begin to execute it in the Microsoft family of Windows OSs.
Therefore, for PE files, their problem-space transformations should be defined within the scope of the format specification requirements, \ie, format-preserving.
For example, one of the transformations that add a new section within the group of Header Information will definitely violate the format of PE files, while adding a new section inside the group of Section Information is acceptable in terms of the format of PE files.
In order to overcome the challenge of format-preserving, one of the straightforward and non-trivial approaches is to carefully check and inspect each candidate transformation according to the formal specification of PE format under the family of Windows OSs.

\subsubsection{Challenge 2: Keep the executability for PE files (\textbf{executability-preserving})}
Although one generated adversarial PE malware is format-preserving, it does not necessarily mean it is executability-preserving as well, which is one of the most challenging properties to be addressed in generating adversarial PE malware.
This is mainly because, the format of PE files only determines its layout and structure under standard specifications, but cannot particularly determine the concrete content of each element inside the layout and structure, and thus cannot guarantee that they can be properly executed.
For example, applying a simple transformation of flipping on byte or even one bit in the ``.data'' section of a given PE file usually does not violate the format specifications of PE files.
However, it is very likely to cause a runtime crash for the transformed PE file as it cannot load the necessary data from the ``.data'' section, thereby preventing its normal execution.

\subsubsection{Challenge 3: Keep the same maliciousness for PE malware (\textbf{maliciousness-preserving})}
Recall that the ultimate goal of adversarial attacks against PE malware detection is to generate practical and realistic adversarial PE malware, which could not only misclassify the target PE malware detection model, but also can keep the same maliciousness as the original PE malware.
However, again one generated adversarial malware that preserves the executability does not necessarily mean it still preserves the sample maliciousness as the original PE malware, \ie, maliciousness-preserving.
Suppose that the generated adversarial malware cannot perform the same maliciousness behaviors (\eg, deleting or encrypting files, modifying registry items, \etc) as the original PE malware, it is totally meaningless and unprofitable for the adversary in the wild.
Therefore, the property of maliciousness-preserving is another significant challenge to be addressed in generating adversarial PE malware.

In short, to address the two aforementioned challenges of \textbf{executability-preserving} and \textbf{maliciousness-preserving}, most of the proposed problem-space attacks claim all transformations adopted are format/executability/maliciousness-preserving, but only with limited inspection and analysis on every specific transformation in either concept or empiric.
Furthermore, due to the complexity of both PE malware and the corresponding executable environments, we argue that it is almost impossible to theoretically prove whether the proposed adversarial attacks can generate successful adversarial PE malware that satisfies the two properties of executability-preserving and maliciousness-preserving.
Alternatively, empirical verification has been reasonably employed in evaluating both executability-preserving and maliciousness-preserving.
In particular, if the generated adversarial PE malware can be properly executed and its runtime status is basically consistent with the runtime status of the original PE malware in the same simulated environments (\eg, sandbox), both executability-preserving and maliciousness-preserving can therefore be empirically and reasonably concluded.
\section{Adversarial Attacks against PE Malware Detection: The State of the Art}\label{sec:adversarial_attacks}

\begin{figure}[htb]
    \centering
    \includegraphics[width=0.95\textwidth,keepaspectratio]{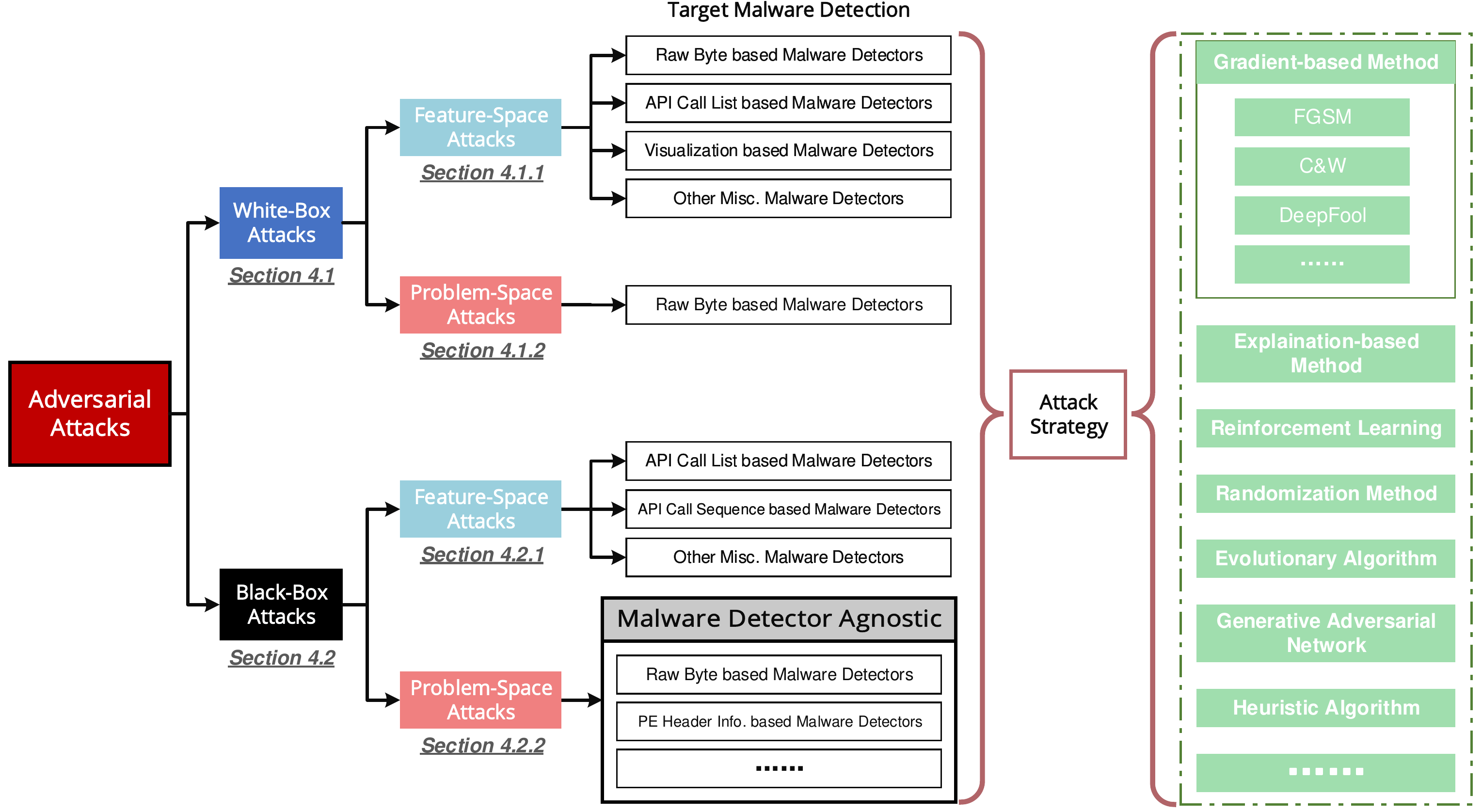}
    \caption{The General Category of Adversarial Attacks against PE Malware Detection.}
    \label{figure:general_category}
\end{figure}

In order to explore the most promising advances of adversarial attacks against PE malware detection, in this section, we comprehensively and systematically categorize state-of-the-art adversarial attacks from different viewpoints, \ie, adversary's knowledge, adversary's space, target malware detection, and attack strategy. 
\Figref{figure:general_category} illustrates the general category of adversarial attacks against PE malware detection of this paper.
In the following subsections, in terms of the adversary's knowledge (\ie, white-box versus black-box), we will first introduce the white-box adversarial attacks against PE malware detection in \Secref{subsec:white_box}, and then introduce the black-box adversarial attacks in \Secref{subsec:black_box}.
Finally, we highlight the summary of state-of-the-art adversarial attacks against PE malware detection in \Secref{subsec:summary_attack}.

\subsection{White-box Adversarial Attacks against PE Malware Detection}\label{subsec:white_box}
Recall that in \Secref{subsec:adversary_attack_concept_taxonomy}, the white-box attack refers to the scenario that the adversary knows full information about the target malware detectors, including architectures, parameters, feature space, the training dataset, \etc
In the following parts, we first further divide the white-box attacks into the \textit{feature-space white-box attacks} (in \Secref{subsubsec:feature_space_white_box}) and the \textit{problem-space white-box attacks} (in \Secref{subsubsec:problem_space_white_box}) based on their corresponding adversary's space (\ie, feature-space versus problem-space).

\subsubsection{Feature-space White-box Attacks against PE Malware Detection}\label{subsubsec:feature_space_white_box}
In this part, we focus on the feature-space white-box attacks against PE malware detection, in which all adversarial manipulations (\eg, adding irrelevant API calls) are performed in the feature space of PE malware (\eg, the API call list).
In particular, to better understand all feature-space white-box attacks in the face of different kinds of PE malware detection models, we group them into the following categories according to the different types of PE malware detection models, including raw byte based malware detectors, API call list based malware detectors, visualization based malware detectors, and other miscellaneous malware detectors.

\textbf{Raw Bytes based Malware Detectors.}
In order to evade the raw bytes based malware detector like MalConv~\cite{raff2017malware}, {Kreuk~\etal}~\cite{kreuk2018deceiving} consider appending or injecting a sequence of bytes, namely the adversarial payload, to the end of PE malware or the slack region, \ie, existing unused continuous bytes of sections in the middle of PE malware.
First, they iteratively generate the adversarial payload with the gradient-based method of FGSM~\cite{goodfellow2014explaining} in the continuous feature space.
Then, to generate the practical adversarial PE malware, they attempt to reconstruct back to the input problem space by directly searching the closest neighbor to the generated adversarial payload.

Similarly, {Suciu~\etal}~\cite{suciu2019exploring} extend the FGSM-based adversarial attacks with two previously proposed strategies (\ie, append-FGSM and slack-FGSM~\cite{kreuk2018deceiving}), and further perform a systematic evaluation to compare the effectiveness of both append and slack strategies against MalConv.
Their experimental results show that slack-FGSM outperforms append-FGSM with a smaller number of modified bytes.
Possible reasons are that the appended bytes of append-FGSM might exceed the maximum size of the model input (\eg, 2MB for MalConv), or that slack-FGSM can make use of surrounding contextual bytes to amplify the power of FGSM since the CNN-based MalConv detector requires the consideration of the contextual bytes within the convolution window.

% {BFA} / {Enhanced-BFA}
% the white-box version in this paper
{Chen~\etal}~\cite{chen2019adversarial} suggest that all those adversarial attacks~\cite{kreuk2018deceiving,suciu2019exploring} append or inject adversarial bytes that are first initialized by random noises and further iteratively optimized, which might lead to inferior attack performance.
To address this issue, {Chen~\etal} propose two novel white-box attacks (\ie, {BFA} and {Enhanced-BFA}) with the saliency vector generated by the Grad-CAM approach~\cite{selvaraju2017grad}.
For {BFA}, it selects the data blocks with significant importance from benign PE files using computed saliency vectors and then appends those data blocks to the end of the original PE malware.
Besides that, Enhanced-BFA is presented to use FGSM to iteratively optimize these perturbations generated by BFA.
Experimental results show that Enhanced-BFA and BFA have comparative attack performances when the number of appending bytes is large, but Enhanced-BFA is ten times more effective than BFA when the number of appending bytes is small.

{Qiao~\etal}~\cite{qiao2022adversarial} propose a white-box adversarial attack against raw bytes based malware detectors like MalConv.
In particular, it first generates a prototype sample to maximize the output of malware detection model towards the target class (\ie, malware) by directly applying the gradient descent algorithm.
Next, to ensure that the generated adversarial PE malware preserves both executability and maliciousness as the input PE malware, it modifies the modifiable part of input PE malware (\ie, bytes between sections, bytes at the end of PE files, and bytes in the newly added section) in a fine-grained manner under the guidance of the generated prototype sample.

\textbf{API Call List based Malware Detectors.}
% $\text{dFGSM}^{k}$ and $\text{rFGSM}^{k}$
% $\text{BGA}^{k}$ and $\text{BCA}^{k}
By flipping the bits of the binary feature vector of malware (``1'' denotes the presence of one Windows API call and ``0'' denotes the absence), {Al-Dujaili~\etal} introduce four kinds of white-box adversarial attacks with $k$ multi-steps, namely $\text{dFGSM}^{k}$, $\text{rFGSM}^{k}$, $\text{BGA}^{k}$, and $\text{BCA}^{k}$~\cite{al2018adversarial}, to attack API call list based malware detectors.
To be specific, $\text{dFGSM}^{k}$ and $\text{rFGSM}^{k}$ are two white-box adversarial attacks that are adapted mainly from the FGSM attack in the continuous feature-space~\cite{goodfellow2014explaining,kurakin2016adversarialA} but extended for the binary feature space via deterministic or randomized rounding, respectively.
$\text{BGA}^{k}$ and $\text{BCA}^{k}$ are two gradient ascent based attacks that update multiple bits or one bit in each step, respectively.

% {GRAMS}
As the winner of ``Robust Malware Detection Challenge''~\cite{advml_2019_challenge} in both attack and defense tracks, {Verwer~\etal}~\cite{verwer2020robust} propose a novel white-box adversarial attack with greedy random accelerated multi-bit search, namely {GRAMS}, which generates functional adversarial API call features and also builds a more robust malware detector in a standard adversarial training setting.
The main idea of {GRAMS} is to perform a greedy search procedure that explores gradient information as the heuristic to indicate which bits to flip among all the binary search space (\ie, 22761 API calls).
At each iteration, {GRAMS} flips $k$ bits of API calls that have the largest absolute gradient and exponentially increases or decreases the value of $k$ depending on whether {GRAMS} finds a better or worse solution.
% sum of the above two
To ensure the functionality of the generated adversarial malware, both~\cite{al2018adversarial} and~\cite{verwer2020robust} limit the attack to flipping `0' to `1', meaning both of them only add irrelevant API calls.

\textbf{Visualization based Malware Detectors.}
% {ATMPA}
Differently, to attack the visualization-based malware detectors, {Liu~\etal}~\cite{liu2019atmpa} propose the first white-box adversarial attack approach, namely Adversarial Texture Malware Perturbation Attack ({ATMPA}), based on adversarial attacks in the domain of image classification tasks~\cite{goodfellow2014explaining, carlini2017towards}.
In particular, {ATMPA} first converts the malware sample to a binary texture grayscale image and then manipulates the corresponding adversarial example with subtle perturbations generated from two existing adversarial attack approaches - FGSM~\cite{goodfellow2014explaining} and C\&W~\cite{carlini2017towards}.
However, the major limitation of {ATMPA} is that the generated adversarial grayscale image of the malware sample destroys the structure of the original malware and thus cannot be executed properly, which makes {ATMPA} unpractical for real-world PE malware detection.

% {COPYCAT}
Similar to {ATMPA}, {Khormali~\etal} present an adversarial attack {COPYCAT}~\cite{khormali2019copycat} against visualization based malware detectors with CNNs.
{COPYCAT} also makes use of existing generic adversarial attacks (\eg, FGSM, PGD, C\&W, MIM, DeepFool, \etc) to generate an adversarial image.
After that, {COPYCAT} appends the generated adversarial image to the end of the original image of malware rather than directly adding it to the original malware image.

% {AMAO}
Differently, to evade visualization based malware detectors, {Park~\etal}~\cite{park2019generation} propose another adversarial attack based on the adversarial malware alignment obfuscation (AMAO) algorithm.
Specifically, a non-executable adversarial image is first generated by the off-the-shelf adversarial attacks in the field of image classification~\cite{goodfellow2014explaining,carlini2017towards}.
Then, in order to attempt to preserve the executability, the adversarial PE malware is finally generated by the AMAO algorithm that minimally inserts semantic $NOP$s at the available insertion points of the original malware such that the modified PE malware is as similar as possible to the generated no-executable adversarial image.

\textbf{Other Miscellaneous Malware Detectors.}
In~\cite{li2020adversarial}, {Li~\etal} first train ML-based malware detection models based on OpCode n-gram features, \ie, the n-gram sequence of operation codes extracted from the disassembled PE file.
Then, the authors employ an interpretation model of SHAP~\cite{NIPS2017_7062} to assign each n-gram feature with an importance value and observe that the 4-gram ``move + and + or + move'' feature is a typical malicious feature as it almost does not appear in the benign PE samples.
Thus, based on this observation, the authors consider a generation method of adversarial PE malware by instruction substitution.
For instance, the ``move + and + or + move'' in 10 sampled malware samples can be replaced with ``push + pop + and + or + push + pop'', which can be used to bypass the malware detectors in their evaluation.

\subsubsection{Problem-space White-box Attacks against PE Malware Detection}\label{subsubsec:problem_space_white_box}
Different from these aforementioned feature-space adversarial attacks that operate in the feature space, there is a growing body of work being proposed to perform problem-space adversarial attacks against PE malware detection.
To be specific, since it is technically feasible to directly modify the raw bytes of PE files with possible constraints, almost all existing problem-space white-box attacks target at raw byte based malware detectors (\eg, MalConv), which are detailed as follows.

% \textbf{Raw Bytes based Malware Detection.}
% {AMB}
In~\cite{kolosnjaji2018adversarial}, {Kolosnjaji~\etal} introduce a gradient-based white-box attack to generate adversarial malware binaries (AMB) against {MalConv}.
To ensure the generated adversarial malware binaries behave identically to the original malware as much as possible, they consider one semantic-preserving manipulation of appending the generated bytes at the end of the original malware file.
The appended bytes are generated by a gradient-descent method of optimizing the appended bytes to maximally increase the probability of the appended PE malware that is predicted as goodware.

To unveil the main characteristics learned by MalConv to discriminate PE malware from benign PE files, {Demetrio~\etal}~\cite{demetrio2019explaining} employ the integrated gradient technique~\cite{sundararajan2017axiomatic} for meaningful explanations and find that MalConv is primarily based on the characteristics learned from PE header rather than malicious content in sections.
Motivated by the observation, they further present a variant gradient-based white-box attack that is almost the same as~\cite{kolosnjaji2018adversarial}.
The only difference is that, {AMB}~\cite{kolosnjaji2018adversarial} injects adversarial bytes at the end of the PE file while this work is limited to changing the bytes inside the specific DOS header in the PE header.

% {RAMEn}
% note there is another black-box version attack of it
In~\cite{demetrio2020adversarial}, {Demetrio~\etal} propose a general adversarial attack framework (RAMEn) against PE malware detectors based on two novel functionality-preserving manipulations, namely \textit{Extend} and \textit{Shift}, which inject adversarial payloads by extending the DOS header and shifting the content of the first section in PE files, respectively.
In fact, the adversarial payload generation can be optimized in both white-box and black-box settings.
For white-box settings, they use the same gradient-based approach as {AMB}~\cite{kolosnjaji2018adversarial} to generate adversarial payloads and then inject payloads via Extend and Shift manipulations.
It is simply noted that, for black-box settings, they use the same genetic algorithm as \cite{demetrio2020efficient} to generate adversarial payloads and then inject them via Extend and Shift manipulations.

% note there is another black box version attack of it
To make it more stealthy than previous adversarial attacks~\cite{kolosnjaji2018adversarial,demetrio2019explaining,demetrio2020adversarial}, {Sharif~\etal} propose a new kind of adversarial attack based on binary diversification techniques which manipulate the instructions of binaries in a fine-grained function level via two kinds of functionality-preserving transformations, \ie, in-place randomization and code displacement~\cite{lucas2021malware}.
In order to guide the transformations that are applied to the PE malware under the white-box setting, they use a gradient ascent optimization to select the transformation only if it shifts its embeddings in a direction similar to the gradient of the attack loss function~\cite{carlini2017towards} with respect to its embeddings.
\afterpage{

\begin{center}
\setlength{\LTcapwidth}{\textwidth}
\renewcommand{\thefootnote}{\fnsymbol{footnote}}
\begin{scriptsize}
\begin{longtable}{>{\raggedleft\arraybackslash}m{19mm}>{\centering\arraybackslash}m{4mm}>{\centering\arraybackslash}m{4mm}>{\centering\arraybackslash}m{3mm}>{\centering\arraybackslash}m{9.5mm}>{\centering\arraybackslash}m{24.5mm}>{\raggedright\arraybackslash}m{23.5mm}>{\raggedright\arraybackslash}m{18mm}>{\centering\arraybackslash}m{3.5mm}>{\centering\arraybackslash}m{4.5mm}>{\centering\arraybackslash}m{4.5mm}}
\caption{Summary of State-of-the-Art Adversarial Attacks against PE Malware Detection. WB/BB is short for the white-box attack and the black-box attack, FS/PS is short for the problem-space attack and the feature-space attack,
% \fullcircle[0.7ex]/\halfcircle[0.7ex]/\emptycircle[0.7ex] denotes fully/partially/emptily preserving the related property.
and \emptycircle[0.7ex]/\fullcircle[0.7ex] denotes emptily/fully preserving the related property.
In particular, as it is almost impossible to theoretically prove both properties of executability-preserving and maliciousness-preserving, we thus use \emptycircle[0.7ex] to denote that related property is preserved neither conceptually nor empirically, \fullcircle[0.7ex] to denote it is preserved both conceptually and empirically, and \halfcircle[0.7ex] to denote it is only preserved conceptually but not empirically without experimental verification.
}

\\
    \toprule
    \endfirsthead
    \multicolumn{11}{l}{Continued Table.}\\
    \toprule
    \endhead
    \bottomrule
    \multicolumn{11}{l}{Table to be continued.}\\
    \endfoot
    \bottomrule
    \endlastfoot

    % \toprule
    \multicolumn{1}{c}{\multirow{2}[4]{19mm}[-15mm]{Attack Names}} & \multirow{2}[4]{4mm}[-15mm]{Year} &  \multirow{2}[0]{*}[-2mm]{\vleftthead{\tabincell{c}{Adversary's Knowledge}}} & \multirow{2}[0]{*}[-5mm]{{\vleftthead{\tabincell{c}{Adversary's Space}}}} & \multicolumn{2}{c}{PE Malware Detection} & \multicolumn{2}{c}{Attack Methods} & \multicolumn{3}{c}{Preservation} \\
    
    \cmidrule(l{1pt}r{1pt}){5-6} \cmidrule(l{1pt}r{1pt}){7-8} \cmidrule(l{1pt}r{1pt}){9-11}  &  &  &  & \multirow{1}[0]{10mm}[-13mm]{Category} & \multicolumn{1}{c}{\multirow{1}[0]{*}[-13mm]{Detection Name}} & \multicolumn{1}{c}{\multirow{1}[0]{*}[-13mm]{Transformation}} & \multicolumn{1}{c}{\multirow{1}[0]{*}[-13mm]{Strategy}} & \vleftthead{Format} & \vlefttheadC{\tabincell{c}{Executability (with\\empirical verification)}} & \vlefttheadC{\tabincell{c}{Maliciousness~(\textit{with}\\\textit{empirical verification})}
    }\\

    \midrule
    \citeauthor{kreuk2018deceiving} \cite{kreuk2018deceiving} & \citeyear{kreuk2018deceiving} & WB & FS & Static & MalConv & Append or inject the adversarial payload & FGSM & \fullcircle & \halfcircle & \halfcircle\\
    \rowcolor{lightgrayfortable}
    \citeauthor{suciu2019exploring} \cite{suciu2019exploring} & \citeyear{suciu2019exploring} & WB & FS & Static & MalConv & Append or inject adversarial payload & FGSM & \fullcircle & \halfcircle & \halfcircle \\
    {BFA}, {Enhanced-BFA} \cite{chen2019adversarial} & \citeyear{chen2019adversarial} & WB & FS & Static & MalConv & Append the selected or optimized bytes from benign PE files  & Grad-CAM or FGSM & \fullcircle & \halfcircle & \halfcircle \\

    \rowcolor{lightgrayfortable}
    {Qiao~\etal} \cite{qiao2022adversarial} & \citeyear{qiao2022adversarial} & WB & FS & Static & MalConv & Append or inject adversarial payload & Gradient-based & \fullcircle & \halfcircle & \halfcircle \\

    {$\text{dFGSM}^{k}$, $\text{rFGSM}^{k}$, $\text{BGA}^{k}$, $\text{BCA}^{k}$ \cite{al2018adversarial}} & \citeyear{al2018adversarial} & WB & FS &  Static & API call list based malware detectors & Add irrelevant API calls & Gradient-based & \fullcircle & \halfcircle & \halfcircle \\
    \rowcolor{lightgrayfortable}
    {GRAMS} \cite{verwer2020robust} & \citeyear{verwer2020robust}  & WB & FS & Static & API call list based malware detectors & Add irrelevant API calls & Gradient-based & \fullcircle & \halfcircle & \halfcircle \\
    
    ATMPA \cite{liu2019atmpa} & \citeyear{liu2019atmpa} & WB & FS & Static & Visualization-based malware detectors & Add adversarial noise to the malware image & FGSM, C\&W & \emptycircle & \emptycircle & \emptycircle \\
    \rowcolor{lightgrayfortable}
    COPYCAT \cite{khormali2019copycat} & \citeyear{khormali2019copycat} & WB & FS & Static & Visualization-based malware detectors & Append adversarial noise generated & FGSM, PGD, C\&W, MIM, DeepFool & \fullcircle & \halfcircle & \halfcircle \\
    AMAO \cite{park2019generation} & \citeyear{park2019generation} & WB & FS & Static & Visualization-based malware detectors & Insert the semantic $NOP$s & FGSM, C\&W & \fullcircle & \halfcircle & \halfcircle \\
    
    \rowcolor{lightgrayfortable}
    \citeauthor{li2020adversarial} \cite{li2020adversarial} & \citeyear{li2020adversarial} & WB & FS & Static & Opcode-based malware detectors & Opcode instruction substitution & Interpretation model SHAP& \fullcircle & \halfcircle & \halfcircle \\
    
    \midrule
    
    AMB \cite{kolosnjaji2018adversarial} & \citeyear{kolosnjaji2018adversarial} & WB & PS & Static & MalConv & Append adversarial bytes & Gradient-based & \fullcircle & \halfcircle & \halfcircle \\
    \rowcolor{lightgrayfortable}
    \citeauthor{demetrio2019explaining} \cite{demetrio2019explaining} & \citeyear{demetrio2019explaining} & WB & PS & Static & MalConv & Modify specific regions in the PE header & Gradient-based & \fullcircle & \halfcircle & \halfcircle \\
    RAMEn \cite{demetrio2020adversarial} & 2020 & WB & PS & Static & MalConv, {Byte-based DNN Model}~\cite{coull2019activation} & DOS Header Extension, Content Shifting & Gradient-based & \fullcircle & \halfcircle & \halfcircle \\
    \rowcolor{lightgrayfortable}
    \citeauthor{lucas2021malware} \cite{lucas2021malware} & \citeyear{lucas2021malware} & WB & PS & Static & MalConv, AvastNet~\cite{krvcal2018deep} & Binary diversification techniques & Gradient-based & \fullcircle & \fullcircle & \fullcircle \\
    
    \midrule
    MalGAN \cite{hu2017generating} & 2017 & BB w/o prob. & FS & Static & API call list based malware detectors & Add irrelevant API calls & GAN & \fullcircle & \halfcircle & \halfcircle \\
    \rowcolor{lightgrayfortable}
    Improved MalGAN \cite{kawai2019improved} & \citeyear{kawai2019improved} & BB w/o prob. & FS & Static & API call list based malware detectors & Add irrelevant API calls & GAN & \fullcircle & \halfcircle & \halfcircle \\
    EvnAttack~\cite{chen2017adversarial} & \citeyear{chen2017adversarial} & BB w prob. & FS & Static & API call list based malware detectors & Add or Remove API calls & Greedy Algorithm & \fullcircle & \emptycircle & \emptycircle \\
    
    \rowcolor{lightgrayfortable}
    \citeauthor{hu2017black} \cite{hu2017black} & 2017 & BB w/o prob. & FS & Dynamic & API call sequence based malware detectors & Insert irrelevant API calls & Generative Model & \fullcircle & \halfcircle & \halfcircle \\
    GADGET \cite{rosenberg2018generic} & \citeyear{rosenberg2018generic} & BB & FS & Dynamic & API call sequence based malware detectors & Insert irrelevant API calls with IAT Hooking & Transferability, Heuristics & \fullcircle & \fullcircle & \halfcircle \\
    \rowcolor{lightgrayfortable}
    ELE \cite{fadadu2019evading} & \citeyear{fadadu2019evading} & BB w prob. & FS & Dynamic & API call sequence based malware detectors & Insert API calls with IAT Hooking & Greedy Algorithm & \fullcircle & \halfcircle & \halfcircle \\
    {BADGER} \cite{rosenberg2020query} & 2020 & BB & FS & Dynamic & API call sequence based malware detectors & Insert API calls with IAT Hooking & Evolutionary Algorithm & \fullcircle & \halfcircle & \halfcircle \\
    
    \rowcolor{lightgrayfortable}
    \citeauthor{rosenberg2020generating} \cite{rosenberg2020generating} & 2020 & BB w/ prob. & FS & Static & EMBER & Predefined modifiable features & Transferability, Explainable ML & \fullcircle & \halfcircle & \halfcircle \\
    SRL \cite{zhang2020semantic} & 2020 & BB w/o prob. & FS & Static & CFG-based malware detectors & Inject semantic $NOP$s into CFG blocks & Reinforcement Learning & \fullcircle & \halfcircle & \halfcircle \\
    
    \midrule
    
    \rowcolor{lightgrayfortable}
    gym-malware \cite{anderson2017evading,anderson2018learning} & 2017 & BB w/o prob. & PS & Static & -- & format-preserving modifications & Reinforcement Learning & \fullcircle & \fullcircle & \halfcircle  \\
    gym-plus \cite{wu2018enhancing} & \citeyear{wu2018enhancing} & BB w/o prob. & PS & Static & -- & format-preserving modifications & Reinforcement Learning & \fullcircle & \halfcircle & \halfcircle  \\
    \rowcolor{lightgrayfortable}
    gym-malware-mini \cite{chen2020generating} & 2020 & BB w/o prob. & PS & Static & -- & format-preserving modifications & Reinforcement Learning & \fullcircle & \halfcircle & \halfcircle \\
    DQEAF \cite{fang2019evading} & \citeyear{fang2019evading} & BB w/o prob. & PS & Static & -- & format-preserving modifications & Reinforcement Learning & \fullcircle & \fullcircle & \fullcircle \\
    \rowcolor{lightgrayfortable}
    RLAttackNet \cite{fang2020deepdetectnet} & \citeyear{fang2020deepdetectnet} & BB w/o prob. & PS & Static & -- & format-preserving modifications & Reinforcement Learning & \fullcircle & \fullcircle & \fullcircle \\
    {AMG-VAC} \cite{ebrahimi2021binary} & \citeyear{ebrahimi2021binary} & BB w/o prob. & PS & Static & -- & format-preserving modifications & Reinforcement Learning & \fullcircle & \halfcircle & \halfcircle \\
    \rowcolor{lightgrayfortable}
    {AIMED-RL} \cite{labaca2021aimed} & \citeyear{labaca2021aimed} & BB w/o prob. & PS & Static & -- & format-preserving modifications & Reinforcement Learning & \fullcircle & \halfcircle & \halfcircle \\
    {AMG-IRL} \cite{li2021irl} & \citeyear{li2021irl} & BB w/o prob. & PS & Static & -- & format-preserving modifications & Reinforcement Learning & \fullcircle & \fullcircle & \fullcircle \\
    
    \rowcolor{lightgrayfortable}
    ARMED \cite{castro2019armed} & 2019 & BB w/ prob. & PS & Static & -- & format-preserving modifications & Randomization & \fullcircle & \fullcircle & \fullcircle \\
    Dropper \cite{ceschin2019shallow} & \citeyear{ceschin2019shallow} & BB w/o prob. & PS & Static & -- & Append strings from goodware \& Packing & Randomization & \fullcircle & \fullcircle & \fullcircle  \\
    \rowcolor{lightgrayfortable}
    \citeauthor{chen2019adversarial} \cite{chen2019adversarial} & \citeyear{chen2019adversarial} & BB w/o prob. & PS & Static & MalConv & Append bytes from benign PE files & Experience-based Randomization & \fullcircle & \halfcircle & \halfcircle \\
    \citeauthor{song2020automatic} \cite{song2020automatic} & \citeyear{song2020automatic} & BB w/o prob. & PS & Static & -- & format-preserving (macro \& micro) modifications & Weighted Randomization & \fullcircle & \fullcircle & \fullcircle  \\
    
    \rowcolor{lightgrayfortable}
    AIMED \cite{castro2019aimed} & 2019 & BB w/ prob. & PS & Static & -- & format-preserving modifications & Genetic Programming & \fullcircle & \fullcircle & \fullcircle \\
    MDEA \cite{wang2020mdea} & \citeyear{wang2020mdea} & BB w/ prob. & PS & Static & MalConv & format-preserving modifications & Genetic Algorithm & \fullcircle & \halfcircle & \halfcircle \\
    \rowcolor{lightgrayfortable}
    GAMMA \cite{demetrio2020efficient} & 2020 & BB w/ prob. & PS & Static & -- & Inject and pad sections from benign PE files & Genetic Algorithm & \fullcircle & \halfcircle & \halfcircle \\
    
    GAPGAN \cite{yuan20black} & \citeyear{yuan20black} & BB w/o prob. & PS & Static & MalConv & Append bytes to the end & GAN & \fullcircle & \halfcircle & \halfcircle \\
    \rowcolor{lightgrayfortable}
    MalFox \cite{zhong2020malfox} & \citeyear{zhong2020malfox} & BB w/o prob. & PS & Static & -- & Obfuscation-like techniques & Convolutional-GAN & \fullcircle & \halfcircle & \halfcircle \\
    
    Targeted occlusion attack \cite{fleshman2018static} & \citeyear{fleshman2018static} & BB w/ prob. & PS & Static & -- & Occlusion of important bytes & Binary Search & \emptycircle & \emptycircle & \emptycircle \\
    \rowcolor{lightgrayfortable}
    \citeauthor{lucas2021malware} \cite{lucas2021malware} & \citeyear{lucas2021malware} & BB w/ prob. & PS & Static & MalConv, AvastNet~\cite{krvcal2018deep} & Binary diversification techniques & Hill-climbing algorithm & \fullcircle & \fullcircle & \fullcircle%
    % \bottomrule
    \label{table:summary_attacks}
\end{longtable}
\end{scriptsize}
\end{center}
}

\subsection{Black-box Adversarial Attacks against PE Malware Detection}\label{subsec:black_box}
Recall that in \Secref{subsec:adversary_attack_concept_taxonomy}, the black-box attack refers to the scenario that the adversary knows nothing about the target PE malware detection models except the outputs, \ie, the malicious/benign label with/without probability.
In the following parts, we first further divide the black-box attacks into the \textit{feature-space black-box attacks} (in \Secref{subsubsec:feature_space_black_box}) and the \textit{problem-space black-box attacks} (in \Secref{subsubsec:problem_space_black_box}) based on their corresponding adversary's space (\ie, feature-space versus problem-space).

\subsubsection{Feature-space Black-box Attacks against PE Malware Detection}\label{subsubsec:feature_space_black_box}
In this part, we focus on the feature-space black-box attacks against PE malware detectors, in which all adversarial operations (\eg, insert irrelevant API calls) are performed in the feature space of PE malware (\eg, the API call sequence) instead of being performed in the PE malware themselves.
As the feature-space attacks rely on the different feature representations from different types of PE malware detectors, we thus group all existing feature-space black-box attacks into the following categories according to different types of PE malware detectors, including API call list based malware detectors, API call sequence based malware detectors, and other miscellaneous malware detectors.

It is worth noting that we distinguish the API call list from the API call sequence for PE malware detectors as follows.
The API call list is a binary feature (\ie, `1' or `0'), indicating whether or not the PE file calls the specific API.
The API call sequence represents the sequence of APIs sequentially that is called by the PE file.
The API call list can be extracted by either static or dynamic analysis techniques while the API call sequence can only be extracted by dynamic analysis techniques.

\textbf{API Call List based Malware Detectors.}
% {MalGAN}
{Hu and Tan}~\cite{hu2017generating} first present a black-box adversarial attack {MalGAN} based on GAN to attack PE malware detectors based on the API call list.
{MalGAN} assumes the adversary knows the complete feature space (\ie, binary features of the 160 system-level API calls) of the target malware detector and considers only adding some irrelevant API calls into the original malware sample for generating the adversarial malware samples in the feature space.
{MalGAN} first builds a differentiable substitute detector to fit the target black-box malware detector and then trains a generator to minimize the malicious probability of generated adversarial malware predicted by the substitute detector.
% {Improved-MalGAN}
Subsequently, {Kawai~\etal}~\cite{kawai2019improved} further present an {Improved-MalGAN} after addressing several issues of {MalGAN} from a realistic viewpoint.
For instance, {Improved-MalGAN} trains the {MalGAN} and the target black-box malware detector with different API call lists while the original {MalGAN} trains with the same API call list.

% {EvnAttack}
In~\cite{chen2017adversarial}, {Chen~\etal} introduce another black-box adversarial attack, namely {EvnAttack}.
{EvnAttack} first employs Max-Relevance~\cite{peng2005feature} to calculate the importance of each API call in classifying PE malware or goodware based on the training set and then ranks those API calls into two sets: $M$ and $B$.
In particular, $M$ contains API calls that are highly relevant to malware, while $B$ contains API calls that are highly relevant to goodware.
Intuitively, {EvnAttack} is a simple and straightforward attack method that manipulates the API call list by either adding the API calls in $B$ or removing the ones in $M$.
Specifically, {EvnAttack} employs a bidirectional selection algorithm that greedily selects API calls for the manipulation of addition or removal based on the fact that how the manipulation influences the loss of the target PE malware detector.

\textbf{API Call Sequence based Malware Detectors.}
% Dynamic analysis based on API call sequences
% RNN-based Attack / Generative RNN
Aiming at attacking RNN-based malware detection models that take the API call sequence as the input, {Hu and Tan}~\cite{hu2017black} propose a generative model based black-box adversarial attack to evade such RNN-based PE malware detectors.
In particular, a generative RNN is trained based on PE malware to generate an irreverent API call sequence that will be inserted into the original API call sequence of the input PE malware, while a substitute RNN model is trained to fit the target RNN-based malware detector based on both benign samples and the gradient information of malware samples from the generative RNN model.

% {GADGET}
% Transferability-based / Jacobian / Gradient
In~\cite{rosenberg2018generic}, {Rosenberg~\etal} propose a generic end-to-end attack framework, namely {GADGET}, against state-of-the-art API call sequence-based malware detectors under black-box settings by the transferability property.
{GADGET} is carried out in three steps:
i) {GADGET} first trains a surrogate model to approximate the decision boundaries of the target malware detector by using the Jacobian-based dataset augmentation method~\cite{papernot2017practical}.
ii) it then performs a white-box attack on the surrogate model to generate the adversarial API call sequence by restricting the insertion of API calls into the original API call sequence.
In more detail, {GADGET} first randomly selects an insert position and then uses a heuristic searching approach to iteratively find and insert the API calls such that the generated adversarial sequence follows the direction indicated by the Jacobian.
iii) to generate practical adversarial malware samples from the adversarial API call sequence, {GADGET} uses a proxy wrapper script to wrap the original malware by calling the additional APIs with valid parameters in the corresponding position based on the generated adversarial API call sequence.

% {ELE}
{Fadadu~\etal}~\cite{fadadu2019evading} propose an executable level evasion (ELE) attack under black-box settings to evade PE malware detectors based on the API call sequence.
The manipulation of {ELE} is restricted only to the addition of new API calls, which are chosen by maximizing the fraction of sub-sequences that have the added API call in the domain of benign samples and minimizing the fraction of sub-sequences that have the added API call in the domain of malware samples.
To further make the modified PE malware can be executed properly, {ELE} uses a novel IAT (\ie, Import Address Table) hooking method to redirect the control in the adversarial code that is attached to the PE malware.
In particular, the adversarial code contains a wrapper function that not only has identical arguments and returns values with the original API function, but also invokes the added API function that is periodically called by the PE malware.

% {BADGER}
Following {GADGET}, {Rosenberg~\etal}~\cite{rosenberg2020query} subsequently propose and implement an end-to-end adversarial attack framework, namely {BADGER}, which consists of a series of query-efficient black-box attacks to misclassify such API call sequence-based malware detector as well as minimize the number of queries.
Basically, to preserve the original functionality, the proposed attacks are limited to only inserting API calls with no effect or an irrelevant effect, \eg, opening a non-existent file.
To solve the problem of which and where the API calls should be inserted in, the authors propose different attacks with or without knowledge of output probability scores, \ie, the score-based attack and the decision-based attack.
For the score-based attack, it uses the self-adaptive uniform mixing evolutionary algorithm~\cite{dang2016self} to optimize the insertion position with the API calls generated by a pre-trained SeqGAN that has been trained to mimic API call sequences of benign samples.
For the decision-based attack, it selects a random insertion position and then inserts the API call with the same position from the pre-trained SeqGAN.
Finally, to make the attacks query-efficient, they first insert a maximum budget of API calls and then employ a logarithmic backtracking method to remove some of the inserted API calls as long as evasion is maintained.

\textbf{Other Miscellaneous Malware Detectors.}
% explainable
% Instead of only adding new features (\eg, add additional API calls, \etc), 
{Rosenberg~\etal}~\cite{rosenberg2020generating} present a transferability-based black-box attack against traditional ML-based malware detection models (\eg, EMBER) by modifying the features instead of just adding new features (\eg, adding API calls) like previous attacks.
The authors first train a substitute model to fit the target black-box malware detector and then use the explainable algorithm to obtain a list of feature importance of the detection result of the original malware on the substitute model.
Subsequently, the authors modify those easily modifiable features with a list of predefined feature values and select a particular value that results in the highest benign probability.

% \textbf{Control Flow Graph based Malware Detection.}
% attacking graph-based malware detection models
% {SRL}
Aiming at attacking graph-based (\ie, CFG) malware detectors~\cite{yan2019classifying}, {Zhang~\etal}~\cite{zhang2020semantic} introduce the first semantic-preserving RL-based black-box adversarial attack named SRL.
To preserve the original functionality of the malware and retain the structure of the corresponding control flow graph, SRL trains a deep RL agent which could iteratively choose basic blocks in CFG and semantic $NOP$s for insertion to modify the PE malware until the generated adversarial malware can successfully bypass the target malware detector.
Their experimental results show that SRL achieves a nearly 100\% attack success rate against two variants of graph neural network based malware detectors.

\subsubsection{Problem-space Black-box Attacks against PE Malware Detection}\label{subsubsec:problem_space_black_box}
In this part, we focus on the problem-space black-box adversarial attacks against PE malware detectors, in which all adversarial operations are performed in the problem space of PE malware under black-box settings, \ie, directly operating the PE malware itself without any consideration of its feature representation (indicating the \textit{problem-space}) as well as the PE malware detectors to be attacked (indicating the \textit{black-box} setting).
It means that, in theory, the problem-space black-box adversarial attacks in the scenario of PE malware detection are completely agnostic to specific PE malware detectors.
Therefore, regardless of any kind of PE malware detectors, we group the problem-space black-box adversarial attacks according to the attack strategies, including reinforcement learning, randomization, the evolutionary algorithm, GAN, and the heuristic algorithm, which are detailed as follows.

\textbf{Reinforcement learning based attacks.}
% {gym-malware}
To expose the weaknesses of current static anti-virus engines, {Anderson~\etal}~\cite{anderson2017evading,anderson2018learning} are the first to study how to automatically manipulate the original PE malware such that the modified PE malware are no longer detected as malicious by the anti-virus engines while do not break the format and functionality.
Particularly, with only knowledge of the binary detection output, they propose a completely black-box adversarial attack based on reinforcement learning (RL), namely {gym-malware}.
The {gym-malware} first defines 10 kinds of format-preserving and functionality-preserving modifications for Windows PE files as the action space available to the agent within the environment.
Then, for any given PE malware, {gym-malware} tries to learn which sequences of modifications in the action space can be used to modify the PE malware, such that the resulting PE malware is most likely to bypass the static anti-virus engines.
Although, {gym-malware} has demonstrated its effectiveness against PE malware detectors, its experimental results also show that RL with an agent of deep Q-network (DQN) or actor-critic with experience replay (ACER)~\cite{sutton2018reinforcement} offers limited improvement compared with the random policy. 
% In theory, {gym-malware} does not break the format and functionality of PE malware, however, it does not conduct any evaluation to verify all the evasive PE malware samples can be executed and preserve the original maliciousness.

On the basis of {gym-malware}, there are multiple follow-up work~\cite{wu2018enhancing,chen2020generating,fang2019evading,fang2020deepdetectnet,ebrahimi2021binary,labaca2021aimed,li2021irl} proposing problem-space black-box adversarial attacks against static PE malware detection models.

% {gym-plus}
In particular, {Wu~\etal}~\cite{wu2018enhancing} propose {gym-plus} based on {gym-malware} with the improvement of adding more format-preserving modifications in the action space and their experimental results show that {gym-plus} with DQN obtains a higher evasion rate than {gym-plus} with the random policy.
% {gym-malware-mini}
Differently, {Chen~\etal}~\cite{chen2020generating} propose {gym-malware-mini} based on {gym-malware} with a limited and smaller action space.
Based on the observation of  most of the format-preserving modifications of {gym-malware} and {gym-plus} are stochastic in nature (\eg, the appending bytes to the new section are chosen at random for simplicity, \etc) and those modifications are not exactly repeatable, {gym-malware-mini} makes 6 kinds of random format-preserving modifications to deterministic modifications, making the RL algorithms easier to learn better policies among limited action space.

% {DQEAF}
Besides that, {Fang~\etal}~\cite{fang2019evading} present a general framework using DQN to evade PE malware detectors, namely {DQEAF}, which is almost identical to gym-malware in methodology except for three implementation improvements as follows.
% Three implementation differences are made based on {gym-malware} as follows. 
1) DQEAF uses a subset of modifications employed in {gym-malware} and guarantees that all of them would not lead to corruption in the modified malware;
2) DQEAF uses a vector with 513 dimensions as the observed state, which is much lower than that in gym-malware;
3) DQEAF makes priority into consideration during the replay of past transitions.
% {RLAttackNet}
{Fang~\etal}~\cite{fang2020deepdetectnet} also observe that the modifications in the action space of {gym-malware} have some randomness and further found that most effective adversarial malware from {gym-malware} are generated by UPX pack/unpacked modifications, which could lead to some training problems with RL due to the non-repeatability of those modifications.
Thus, they first reduce the action space to 6 categories having certain deterministic parameters and then propose an improved black-box adversarial attack, namely {RLAttackNet}, based on the {gym-malware} implementation.

% {AMG-VAC}
{Ebrahimi~\etal}~\cite{ebrahimi2021binary} suggest that the RL-based adversarial attacks against PE malware detectors normally employ actor-critic or DQN, which are limited in handling environments with combinatorially large state space.
Naturally, they propose an improved RL-based adversarial attack framework of {AMG-VAC} on the basis of {gym-malware}~\cite{anderson2017evading, anderson2018learning} by adopting the variational actor-critic, which has been demonstrated to be the state-of-the-art performance in handling environments with combinatorially large state space.
% {AIMED-RL}
As previous RL-based adversarial attacks tend to generate homogeneous and long sequences of transformations, {Labaca-Castro~\etal}~\cite{labaca2021aimed} thus present an RL-based adversarial attack framework of {AIMED-RL} as well.
The main difference between {AIMED-RL} and other RL-based adversarial attacks is that {AIMED-RL} introduces a novel penalization to the reward function for increasing the diversity of the generated sequences of transformations while minimizing the corresponding lengths.
% {AMG-IRL}
{Li and Li}~\cite{li2021irl} suggest that existing RL-based adversarial attacks~\cite{anderson2017evading,anderson2018learning,fang2019evading} employ the artificially defined instant reward function and environment, which are highly subjective and empirical, potentially leading to non-converge of the RL algorithm.
Therefore, in order to address the issue of the subjective and empirical reward function, they present an Inverse RL-based adversarial malware generation method, namely {AMG-IRL}, which could autonomously generate the flexible reward function according to the current optimal strategy.

% comparison
In short, compared to~\cite{wu2018enhancing,chen2020generating,ebrahimi2021binary,labaca2021aimed}, the adversarial malware samples generated by~\cite{fang2019evading,fang2020deepdetectnet,li2021irl} are verified not only for executability within the Cuckoo sandbox~\cite{cuckoo2020}, but also verified for the original maliciousness via comparing the function call graph between the before and after malware samples with IDA Pro~\cite{idapro2020}.

\textbf{Randomization based attacks.}
% {ARMED}
To fully automatize the process of generating adversarial malware without corrupting the malware functionality under the black-box setting, {Castro~\etal}~\cite{castro2019armed} propose {ARMED} -- automatic random malware modifications to evade detection.
{ARMED} first generates the adversarial PE malware by randomly applying manipulations among 9 kinds of format-preserving modifications from~\cite{anderson2017evading,anderson2018learning}, and then employs the Cuckoo sandbox to test the functionality of the generated adversarial malware samples.
In case the functionality test fails, the above steps would be re-start with a new round until the functionality test successes.

% {Dropper}
{Ceschin~\etal}~\cite{ceschin2019shallow} find that packing the original PE malware with a distinct packer~\cite{telock_2020, cheng2018towards} or embedding the PE malware in a dropper~\cite{dropper_2020, kwon2015dropper} is an effective approach in bypassing ML-based malware detectors when combined with appending goodware strings to malware.
However, some of the generated adversarial PE malware suffer from either not being executed properly or being too large in size.
To solve the challenges, the authors implemented a lightweight dropper, namely Dropper, which first creates an entire new PE file to host the original PE malware and then randomly chooses goodware strings to be appended at the end of the newly created PE file.

% black-box attack version
Similar to the white-box attack of {Enhanced-BFA}, {Chen~\etal}~\cite{chen2019adversarial} also introduce another black-box version of adversarial attack against {MalConv}.
First, it continuously selects data blocks at random from goodware and appends them to PE malware to generate adversarial PE malware.
After performing multiple random attacks as above, it then calculates the contribution degree of each data block based on the experience of successful trajectories of data blocks.
Finally, it appends the data blocks to the end of PE malware according to the order of their contribution degrees.

% {macro/micro-actions minimizer}
In~\cite{song2020automatic}, {Song~\etal} propose an automatic black-box attack framework that applies a sequence of actions to rewrite PE malware for evading PE malware detectors.
In particular, to generate adversarial malware samples with a minimal set of required actions from macro/micro actions, the authors employ an action sequence minimizer that consists of three steps.
First, it randomly selects macro-actions according to the previously updated weights of actions as the action sequence to rewrite the original malware.
Second, it tries to remove some unnecessary macro-actions from the action sequence to generate a minimized adversarial malware, and increases the weights of effective actions for updating.
Finally, for every macro-action in the minimized adversarial malware, it attempts to replace the macro-action with a corresponding micro-action.
Besides that, the proposed framework can also help explain which features are responsible for evasion as every required action in adversarial malware samples corresponds to a type of affected feature.

\textbf{Evolutionary algorithm based attacks.}
% {AIMED}
Following the black-box adversarial attack framework of {ARMED}~\cite{castro2019armed}, {Castro~\etal}~\cite{castro2019aimed} propose AIMED, which employs a genetic programming approach rather than randomization methods to automatically find optimized modifications for generating adversarial PE malware.
Firstly, 9 kinds of format-preserving modifications are introduced and applied to the original PE malware to create the population, and then each modified sample in the population is evaluated by a fitness function in terms of functionality, detection ratio, similarity, and the current number of generations.
Secondly, if the modified PE malware fails to bypass the PE malware detector, AIMED implements the classic genetic operations, including selection, crossover, and mutation, which are repeated until all generated adversarial PE malware can evade the PE malware detector.
Experiments demonstrate the time of generating the same number of successful adversarial malware is reduced up to 50\% compared to the previous random based approach~\cite{castro2019armed}.

% {MDEA}
Similarly, {Wang and Miikkulainen}~\cite{wang2020mdea} propose to retrain MalConv with the adversarial PE malware, namely {MDEA}.
To generate the adversarial malware samples, {MDEA} adjusts 10 kinds of format-preserving manipulations from~\cite{anderson2017evading,anderson2018learning} as the action space and employs a genetic algorithm to optimize different action sequences by selecting manipulations from the action space until the generated adversarial malware bypasses the target malware detectors.
In particular, as some manipulations are stochastic in nature, {MDEA} limits each manipulation with a parameter set to make the adversarially trained models converge within an acceptable time.
However, the generated adversarial malware samples by {MDEA} are not tested for functionality like {AIMED} and {ARMED}.

% {GAMMA}
To omit the functionality testing (\eg, sandbox required) and further speed up the computation of prior work~\cite{castro2019aimed,castro2019armed}, {Demetrio~\etal} introduced an efficient black-box attack framework, namely GAMMA~\cite{demetrio2020efficient}.
For GAMMA, the generation approaches of adversarial malware are limited to two types of functionality-preserving manipulations: section injection and padding.
Specifically, benign contents are extracted from the goodware as adversarial payloads to inject either into some newly-created sections (section injection) or at the end of the file (padding).
The main idea of GAMMA is to formalize the attack as a constrained minimization problem which not only optimizes the probability of evading detection, but also penalizes the size of the injected adversarial payload as a regularization term.
To solve the constrained minimization problem, GAMMA also employs a genetic algorithm, including selection, crossover, and mutation, to generate adversarial PE malware to bypass the malware detector with few queries as well as small adversarial payloads.

\textbf{GAN based attacks.}
% {GAPGAN}
In~\cite{yuan20black}, {Yuan~\etal} present {GAPGAN}, a novel GAN-based black-box adversarial attack framework that is performed at the byte-level against DL-based malware detection, \ie, MalConv.
Specifically, {GAPGAN} first trains a generator and a discriminator concurrently, where the generator intends to generate adversarial payloads that would be appended at the end of original malware samples, and the discriminator attempts to imitate the black-box PE malware detector to recognize both the original PE goodware and the generated adversarial PE malware.
After that, {GAPGAN} uses the trained generator to generate the adversarial payload for every input PE malware and then appends the adversarial payload to the corresponding input PE malware, which generates the adversarial PE malware while preserving its original functionalities.
Experiments show that {GAPGAN} can achieve a 100\% attack success rate against MalConv with only appending payloads of 2.5\% of the total length of the input malware samples.

% {MalFox}
{Zhong~\etal}~\cite{zhong2020malfox} propose a convolutional generative adversarial network-based (C-GAN) framework, namely {MalFox}, which can generate functionally indistinguishable adversarial malware samples against realistic black-box antivirus products by perturbing malware files based on packing-related methods.
In general, MalFox consists of 5 components: PE Parser, Generator, PE Editor, Detector, and Discriminator, where Detector is the target black-box malware detector (\eg, antivirus product) and Discriminator is an API call-based malware detector, representing the discrimination model in GAN.
First, the API call list (\ie, DLL and system functions) of the original malware sample is extracted as a binary feature vector by PE Parser;
Second, the Generator takes both the malware feature vector and a sampled 3-dimensional Gaussian noise as input to produce a 3-dimensional perturbation path, indicating whether each of the three perturbation methods (\ie, Obfusmal, Stealmal, and Hollowmal) is adopted.
Third, following the produced perturbation path, the PE editor generates the adversarial malware sample with corresponding perturbation methods.
Finally, the Generator will stop training until the generated adversarial PE malware fails to be recognized by Discriminator.

\textbf{Heuristic based attacks.}
Aiming to quantify the robustness of PE malware detectors ranging from two ML-based models to four commercial anti-virus engines, {Fleshman~\etal}~\cite{fleshman2018static} propose one novel targeted occlusion black-box attack for comparing with three pre-existing evasion techniques, \ie, random-based gym-malware~\cite{anderson2017evading, anderson2018learning}, obfuscation through packing~\cite{cheng2018towards,aghakhani2020malware}, and malicious ROP injection~\cite{poulios2015ropinjector}.
For the proposed targeted occlusion attack, it first uses the occlusion binary search method to identify the most important byte region according to the changes of the malicious probability for a given PE malware detector, and then replaces the identified region with completely random bytes or a contiguous byte region selected randomly from benign samples.
However, we believe that adversarial malware samples generated by the targeted occlusion attack are destructive because the replacement could prevent the generated malware from being executed, not to mention maintain the original maliciousness.

% note there is another white box version attack of it
Based on the two kinds of functionality-preserving transformations (\ie, in-place randomization and code displacement) that manipulate the instructions of binaries in a fine-grained function level, {Lucas~\etal}~\cite{lucas2021malware} also propose another black-box version of adversarial attack based on a general hill-climbing algorithm.
This black-box attack is basically similar to the white-box version and the only difference is how the attempted transformation is selected.
Specifically, the black-box attack first queries the model after attempting to apply one of the transformations to the PE malware, and then accepts the transformation only if the corresponding benign probability increases.

\subsection{Summary of Adversarial Attacks against PE Malware Detection}\label{subsec:summary_attack}
In the research field of PE malware detection, adversarial attack methods have been rapidly proposed and developed in recent years since 2017.
For all four adversarial attack categories (\ie, feature-space white-box, problem-space white-box, feature-space black-box, and problem-space black-box) detailed above, their generated adversarial PE malware is becoming more and more practical and effective in attacking the target PE malware detection.
We summarize the state-of-the-art adversarial attacks against PE malware detection in \Tabref{table:summary_attacks}, which demonstrates the adversarial attack methods and their corresponding categories and characteristics in detail.

For all white-box attacks against PE malware detection, regardless of the adversary's space (feature-space or problem-space), it is clearly observed from \Tabref{table:summary_attacks} that almost all of the white-box attacks adopt the optimization of gradient-based methods as their attack strategies.
Actually, gradient-based optimization and its variants have been widely adopted in the domain of adversarial attacks against image classification models~\cite{goodfellow2014explaining,carlini2017towards}.
However, it is infeasible and impractical to directly apply the gradient-based optimization methods to generate ``realistic'' adversarial PE malware due to the problem-feature space dilemma~\cite{quiring2019misleading}.
Therefore, it is primarily important to adapt the existing gradient-based methods (\eg, FGSM, C\&W) within constraints of feasible transformations (\eg, append adversarial bytes, add irrelevant API calls) according to the different types of PE malware detectors, indicating the white-box attacks normally depend on specific malware detectors.

Compared with the white-box attacks, the black-box attacks are more realistic and practical in the wild due to their minimal reliance on knowledge about the target malware detector.
For the feature-space black-box attacks, as they are actually performed in the feature space of PE files, existing adversarial attack methods generally devise corresponding feasible transformations (\eg, add irrelevant API calls) for PE malware detectors with different feature spaces (\eg, API call list based malware detectors), indicating the black-box attacks are normally malware detector specific.
However, for the problem-space black-box attacks with the most strict requirements due to their manipulations in the problem-space, most of them are malware detector agnostic, meaning that these adversarial attack methods can be used to attack any kind of PE malware detectors in theory.

In terms of property preservation (\ie, format, executability, and maliciousness), for all kinds of adversarial attacks against PE malware detection, it is also observed from \Tabref{table:summary_attacks} that most of them can only guarantee the format-preserving rather than executability-preserving and maliciousness-preserving.
In particular, several adversarial attack methods (\eg, ATMPA~\cite{liu2019atmpa}, COPYCAT~\cite{khormali2019copycat} and the target occlusion attack~\cite{fleshman2018static}) might destroy the fixed layout and grammar of the PE format that is necessary to load and execute the PE file.
On the other hand, for those adversarial attacks like~\cite{lucas2021malware,fang2019evading,fang2020deepdetectnet,li2021irl,castro2019armed,ceschin2019shallow,song2020automatic,castro2019aimed}, they are verified not only for the executability, but also verified experimentally whether the generated adversarial PE malware keeps the same maliciousness as the original PE malware, which is strongly recommended and advocated in our opinion.
\section{Adversarial Defenses for PE Malware Detection}\label{sec:adversarial_defenses}

As various adversarial attacks continue to be proposed and evaluated, adversarial defense methods are meanwhile proposed accordingly.
In fact, the rapid development of adversarial attacks and counterpart defenses constitutes a constant arms race, meaning a new adversarial attack can easily inspire the defender to devise a novel corresponding defense, and a newly proposed defense method will inevitably lead the attacker to design a new adversarial attack for profit.
Therefore, it is important to explore the most promising advances of adversarial defenses for Windows PE malware detection against adversarial attacks.
Although there are currently few researchers that specifically and exclusively propose adversarial defenses for Windows PE malware detection, most of the aforementioned research efforts on adversarial attacks might more or less present corresponding defense methods.
In this section, we summarize the state-of-the-art adversarial defenses for PE malware detection in recent years, mainly including adversarial training and several other defenses.

\subsection{Adversarial Training}
Adversarial training is one of the mainstream adversarial defenses to resist adversarial attacks regardless of specific applications, \eg, image classification, natural languages processing, speech recognition, \etc
Intuitively, adversarial training refers to the defense mechanism that attempts to improve the robustness of the ML/DL model by re-training it with the generated adversarial examples with/without original training examples.
In the scenario of defending against adversarial attacks for PE malware detection, adversarial training and its variants are also widely adopted and we detail them as follows.

Most studies on adversarial defenses~\cite{hu2017generating,anderson2018learning,chen2017adversarial,wu2018enhancing,chen2019adversarial,wang2020mdea,zhang2020semantic} based on adversarial training follow a similar procedure, in which adversarial PE malware or adversarial PE features are first generated and then re-train a corresponding ML/DL model based on the adversarial PE malware/features with/without the original PE malware/features.
For instance, {Hu and Tan} re-train a random-forest-based malware detector based on the original API calls as well as the adversarial API calls generated by the adversarial attack of MalGAN~\cite{hu2017generating}.
{Anderson~\etal}~\cite{anderson2018learning} first exploit the gym-malware to generate adversarial PE malware, and then re-train the malware detection model of EMBER based on the original PE files and the generated adversarial PE malware.
%%%%%%%%%
Differently, in addition to adversarial training with adversarial API calls generated by EvnAttack, {Chen~\etal}~\cite{chen2017adversarial} also present a new secure-learning framework for PE malware detection, namely {SecDefender}, which adds a security regularization term by considering the evasion cost of feature manipulations by attackers.

To sum up, for those adversarial defenses based on adversarial training, it is generally observed that, 1) adversarial training is experimentally demonstrated to mitigate one or several adversarial attacks to some extent; 2) adversarial training inevitably introduces significant additional costs in generating adversarial examples during the training process.

\subsection{Other Defense Methods}

As the first place in the defender challenge of the Microsoft's 2020 Machine Learning Security Evasion Competition~\cite{mlsec_attack}, {Quiring~\etal} present a combinatorial framework of adversarial defenses -- {PEberus}~\cite{quiring2020against} based on the following three defense methods as follows.
\begin{enumerate}
    \item A PE file is passed into the semantic gap detectors, which are used to check whether the PE file is maliciously challenged based on three simple heuristics, \ie, slack space, overlay, and duplicate.
    For instance, a considerably high ratio of the overlay to the overall size usually indicates the PE file might have appended bytes to the overlay.
    \item If the PE file is not detected by the semantic gap detectors, PEberus employs the ensemble of existing malware detectors (\eg, EMBER~\cite{anderson2018ember}, the monotonic skip-gram model~\cite{incer2018adversarially}, and the signature-based model~\cite{yara2020}) and use the max voting for predictions.
    \item PEberus also employs a stateful nearest-neighbor detector~\cite{chen2020stateful} which continuously checks if the PE file is similar to any of the previously detected malware in the history buffer.
\end{enumerate}
In short, with PEberus, a PE file is predicted as malicious if any of the above three defense methods predicts it as malicious.

{Lucas~\etal}~\cite{lucas2021malware} attempt to mitigate their proposed adversarial attacks by exploiting two defensive mechanisms, \ie, binary normalization and instruction masking.
For the defense of binary normalization, it applies the transformation of in-place randomization iteratively into the PE file to reduce its lexicographic representation, so that its potential adversarial manipulations can be undone before inputting it into the downstream PE malware detector.
Similarly, the instruction masking defense first selects a random subset of the bytes that pertain to instructions and then masks it with zeros before inputting the PE file into the PE malware detector.
\section{Discussions}\label{sec:discussion}

The previous sections of \Secref{sec:adversarial_attacks} and \Secref{sec:adversarial_defenses} enable interested readers to have a better and faster understanding with regard to the adversarial attacks and defenses for Windows PE malware detection.
In the following subsections, we first present the other related attacks against PE malware detection beyond the aforementioned adversarial attacks in ~\Secref{subsec:beyond_adversarial_attacks} and then shed some light on research directions as well as opportunities for future work in \Secref{subsec:future_direction}.

\subsection{Beyond Adversarial Attacks}\label{subsec:beyond_adversarial_attacks}

\subsubsection{Universal Adversarial Perturbation}
Universal Adversarial Perturbation (UAP) is one special type of adversarial attack in which an identical single perturbation can be applied over a large set of inputs for misclassifying the target model in the testing phase.
In order to generate the problem-space UAP against PE malware classifiers in the wild, {Labaca-Castro~\etal}~\cite{labaca2021universal} first prepare a set of available transformations (\eg, adding sections to the PE file, renaming sections, pack/unpacking, \etc) for Windows PE files, and then perform a greedy search approach to identify a short sequence of transformations as the UAP for the PE malware classifier.
In particular, if the identified short sequence of transformations representing the UAP is applied to any PE malware, then the resulting adversarial PE malware can evade the target PE malware classifier with high probability while preserving the format, executability, and maliciousness.

\subsubsection{Training-time Poisoning Attacks}
Different from the adversarial attacks that are performed in the testing phase, the poisoning attacks aim to manipulate the training phase, such that the resulting poisoned model $f_b$ has the same prediction on a clean set of inputs as the cleanly trained model $f_c$ but has an adversarially-chosen prediction on the poisoned input (\ie, the input associated a specific backdoor trigger).

Aiming at misclassifying a specific family of malware as goodware, {Sasaki~\etal}\cite{sasaki2019embedding} propose the first poisoning attack against ML-based malware detectors.
They assume the strictest attack scenario that the adversary not only has full knowledge of the training dataset and the learning algorithm, but also can manipulate the training samples (\ie, malware) in the feature space as well as their labels.
In particular, the poisoning attack framework first selects one specific malware family, and then utilizes the same optimization algorithm like~\cite{munoz2017towards} to generate the poisoning samples in the feature-space, which are finally adopted to train the poisoned model.

However, poisoning one entire family of malware is much easier to be detected if the defender checks all malware families individually.
In contrast, poisoning one specific instance of malware is a more challenging problem, by which any malware associated with a backdoor trigger would be misclassified as goodware, regardless of their malware families.
In addition, it is almost impossible to control and manipulate the labeling process for the poisoning data samples for the adversary, since most of the training samples normally are labeled with multiple independent anti-virus engines by security companies and further used to train the malware detection models.
Therefore, considering the practicality of the attack scenario in the wild, both following \cite{severi2021explanation} and \cite{shapira2020being} belong to the clean-label poisoning attack~\cite{shafahi2018poison}, in which the adversary can control and manipulate the poison instance itself, but cannot control the labeling of the poison instance.

In~\cite{severi2021explanation}, targeting the feature-based ML malware classifier, {Severi~\etal} consider the scenario where the adversary can know and manipulate the feature-space of software to build the poisoned goodware (\ie, goodware with the backdoor trigger), some of which can be obtained by security companies to train the feature-based ML malware classifier.
To be specific, the backdoor trigger is created by presenting a greedy algorithm to select coherent combinations of relevant features and values based on the explainable machine learning technique (\ie, SHAP~\cite{NIPS2017_7062}).
Experimental results indicate that the case of 1\% poison rate and 17 manipulated features results in an attack success rate of about 20\%.

{Shapira~\etal}~\cite{shapira2020being} argue that the attack assumption of feature-space manipulations in~\cite{severi2021explanation} is unrealistic and unreasonable for real-world malware classifiers.
In pursuit of a poisoning attack in a problem space, {Shapira~\etal} propose a novel instance poisoning attack by first selecting the goodware that is most similar to the target malware instance and then adding sections to the goodware for adversarially training the poisoned model.
Actually, the manipulation of adding sections acts as the backdoor trigger, which can remain the functionality of the associated goodware as well as the malware instance.
During the testing phase, the target malware instance associated with the backdoor trigger will be misclassified as benign by the poisoned model.

\subsubsection{Model Steal Attacks}
In order to approximate (\ie, steal) a remote deployed detection model $f_b$ of PE malware under a strictly black-box setting, {Ali and Eshete}~\cite{ali2020best} propose a best-effort adversarial approximation method, which mainly relies on limited training samples and publicly accessible pre-trained models.
The proposed best-effort adversarial approximation method leverages feature representation mapping (\ie, transforming the raw bytes of each PE to an image representation) and cross-domain transferability (\ie, taking advantage of pre-trained models from image classification) to approximate the PE malware detection model $f_b$ by locally training a substitute PE malware detection model $f_s$.
Experimental results show that the approximated model $f_s$ is nearly 90\% similar to the black-box model $f_b$ in predicting the same input samples.

\subsection{Future Directions and Opportunities}\label{subsec:future_direction}

\subsubsection{Strong Demands for Robust PE Malware Detection}
As described in \Secref{subsection:learning_framework_for_PE_malware_detection} and \Secref{sec:adversarial_defenses}, although there are a large number of detection methods for PE malware, there are only very limited adversarial defense methods for building more robust PE malware detection models against adversarial attacks.
In general, almost all current adversarial defense methods are empirical defenses based on various heuristics (\eg, adversarial training, input transformation, \etc), which are usually only effective for one or a few adversarial attack methods, indicating these empirical defenses are normally attack-specific.
On the one hand, with the massive existence and continuous evolution of PE malware and corresponding adversarial malware, we argue that the demand for empirical defense methods against them is also likely to rise accordingly to build more robust PE malware detection models.
On the other hand, the substantial work of \textit{robustness certification} applied in image classification tasks suggests a strong demand for PE malware detection models with theoretically guaranteed robustness, as there is no work studying the certified defenses against various adversarial attacks until now.
We argue that certifying the robustness of malware detection models not only helps users comprehensively evaluate their effectiveness under any attack, but also increases the transparency of their pricing in cloud services (\eg, malware detection as a service).

\subsubsection{Practical and Efficient Adversarial Attacks against Commercial Anti-viruses in the Wild}
As introduced and summarized in \Secref{sec:adversarial_attacks} and \Tabref{table:summary_attacks}, the current adversarial attack methods against PE malware detection have been devoted to developing problem-space black-box adversarial attacks, which usually take a similar and typical attack procedure. In general, the attack procedure first defines a set of available transformations (\eg, inject adversarial payload, insert the semantic $NOP$s, \etc) in the problem-space, and then employs a variety of search strategies (\eg, gradient-based, reinforcement learning, genetic algorithm, \etc) to choose a sequence of transformations which can be applied to the original PE malware for generating the adversarial PE malware.
Based on the above observation, we argue there is still much room for improving both the effectiveness and efficiency of adversarial attacks against PE malware detection in two aspects: 
i) devising and defining more practical and stealthier transformations for PE malware.
For example, instead of simply inserting the $NOP$s in the blocks of CFGs~\cite{zhang2020semantic} that is easily noticed and removed by defenders, the transformation of splitting one block of CFGs into multiple iteratively called blocks is much stealthier to be noticed and removed.
ii) designing and implementing more efficient search strategies to accelerate the generation of adversarial PE malware.
We argue that it is quite time-consuming for both RL and genetic algorithm based search strategies.

In addition, it is clearly observed that most existing adversarial attacks target PE malware detection based on static analysis rather than dynamic analysis, which is particularly unknown for both attackers and defenders. 
However, the mainstream commercial anti-virus software/service used by end users of laptops and servers normally employs a hybrid defense solution with both static analysis and dynamic analysis.
Therefore, it is extremely important and urgently demanded to devise and implement practical and efficient adversarial attacks against PE malware detection based on dynamic analysis and commercial anti-viruses.

\subsubsection{Lack of Benchmark Platforms for Research}
As the continuous emergence of adversarial attacks against Windows PE malware detection has led to the constant arms race between adversarial attacks and defense methods, it is challenging to quantitatively understand the strengths and limitations of these methods due to incomplete and biased evaluation.
In fact, in terms of other research fields like adversarial attacks in images, texts, and graphs, there are a large number of corresponding toolboxes and platforms~\cite{papernot2016technical_cleverhans,ling2019deepsec,zeng2020openattack,li2020deeprobust} that have been implemented and open-sourced.
Based on these toolboxes and platforms, subsequent practitioners and researchers can not only exploit them to evaluate the actual effectiveness of previously proposed methods, but also take them as a cornerstone to implement their own proposed attacks and defenses, thereby reducing the time consumption for repetitive implementations.
Therefore, in order to further advance the research on adversarial attacks against Windows PE malware detection, we argue that it is significantly important to design and implement a benchmark platform with a set of benchmark datasets, representative PE malware detection to be targeted, state-of-the-art adversarial attack \& defense methods, performance metrics, and environments for a comprehensive and fair evaluation.
\section{Conclusion}\label{sec:conclusion}

In this paper, we conduct a comprehensive review on the state-of-the-art adversarial attacks against Windows PE malware detection as well as the counterpart adversarial defenses.
To the best of our knowledge, this is the first work that not only manifests the unique challenges of adversarial attacks in the context of Windows PE malware in the wild, but also systematically categorizes the extensive work from different viewpoints in this research field.
Specifically, by comparing the inherent differences between PE malware and images that have been explored originally and traditionally, we present three unique challenges (\ie, format-preserving, executability-preserving, and maliciousness-preserving) in maintaining the semantics of adversarial PE malware for practical and realistic adversarial attacks.
Besides, we review recent research efforts in both adversarial attacks and defenses, and further develop reasonable taxonomy schemes to organize and summarize the existing literature, aiming at making them more concise and understandable for interested readers.
Moreover, beyond the aforementioned adversarial attacks, we discuss other types of attacks against Windows PE malware detection and shed some light on future directions.
Hopefully, this paper can serve as a useful guideline and give related researchers/practitioners a comprehensive and systematical understanding of the fundamental issues of adversarial attacks against PE malware detection, thereby becoming a starting point to advance this research field.

%%       \citet{<label>} ==> Jones et al. [21]
%%       \citep{<label>} ==> [21]
%%

\section*{Acknowledgement}
This project is supported by 
the Strategic Priority Research Program of the Chinese Academy of Sciences under Grant No.XDA0320000,
the National Natural Science Foundation of China under No.62202457 and No.U1936215,
and the project funded by China Postdoctoral Science Foundation under No.2022M713253.
Yaguan Qian is also supported by the Key Program of Zhejiang Provincial Natural Science Foundation of China under No.LZ22F020007.

\bibliographystyle{elsarticle-num-names} 
\bibliography{ReferenceMain,ReferenceDetection}

\end{document}